\documentclass[acmlarge,screen]{acmart}
\AtBeginDocument{%
  \providecommand\BibTeX{{%
    \normalfont B\kern-0.5em{\scshape i\kern-0.25em b}\kern-0.8em\TeX}}}

\setcopyright{rightsretained}
\acmJournal{IMWUT}
\acmYear{2022} \acmVolume{6} \acmNumber{3} \acmArticle{137} \acmMonth{9} \acmPrice{}\acmDOI{10.1145/3550326}

\usepackage{caption}
\usepackage{subcaption}
\usepackage{booktabs,amsfonts,dcolumn}
\usepackage{multirow}
\usepackage{fancyvrb,cprotect}
\begin{document}
%\nolinenumbers
\title{Psychophysiological Arousal in Young Children Who Stutter: An Interpretable AI Approach}

\author{\href{https://orcid.org/0000-0002-7016-6220}{Harshit Sharma}}
\affiliation{%
  \institution{Syracuse University}
  \city{Syracuse}
  \country{USA}}
\email{hsharm04@syr.edu}

\author{\href{https://orcid.org/0000-0002-5261-5440}{Yi Xiao}}
\affiliation{%
  \institution{Syracuse University}
  \city{Syracuse}
  \country{USA}}
\email{yxiao54@syr.edu}

\author{\href{https://orcid.org/0000-0002-4216-683X}{Victoria Tumanova}}
\affiliation{%
  \institution{Syracuse University}
  \city{Syracuse}
  \country{USA}}
\email{vtumanov@syr.edu}

\author{\href{https://orcid.org/0000-0002-0807-8967}{Asif Salekin}}
\thanks{Asif Salekin is the corresponding author.}
\affiliation{%
  \institution{Syracuse University}
  \city{Syracuse}
  \country{USA}}
\email{asalekin@syr.edu}
\renewcommand{\shortauthors}{Sharma et al}


\begin{abstract}
The presented first-of-its-kind study effectively identifies and visualizes the second-by-second pattern differences in the physiological arousal of preschool-age children who do stutter (CWS) and who do not stutter (CWNS) while speaking perceptually fluently in two challenging conditions: speaking in stressful situations and narration. The first condition may affect children’s speech due to high arousal; the latter introduces linguistic, cognitive, and communicative demands on speakers. We collected physiological parameters data from 70 children in the two target conditions. First, we adopt a novel modality-wise multiple-instance-learning (MI-MIL) approach to classify CWS vs. CWNS in different conditions effectively. The evaluation of this classifier addresses four critical research questions that align with state-of-the-art speech science studies' interests. Later, we leverage SHAP classifier interpretations to visualize the salient, fine-grain, and temporal physiological parameters unique to CWS at the population/group-level and personalized-level. While group-level identification of distinct patterns would enhance our understanding of stuttering etiology and development, the personalized-level identification would enable remote, continuous, and real-time assessment of stuttering children's physiological arousal, which may lead to personalized, just-in-time interventions, resulting in an improvement in speech fluency. The presented MI-MIL approach is novel, generalizable to different domains, and real-time executable. Finally, comprehensive evaluations are done on multiple datasets, presented framework, and several baselines that identified notable insights on CWSs’ physiological arousal during speech production.
\end{abstract}


% this study is the first to understand second by second temporal patterns from group and personalized level.


\maketitle
\section{Introduction}\label{introduction-section}

The recent advancement in technology has resulted in the production of state-of-the-art sensors which provide an accurate reading of various physiological signals with minimum intrusion and only pose minor limitations in a person’s mobility. The physiological data collected by these sensors give insight into the human affective states and allow us to examine how emotion influences human thought and behavior. \textit{The goal of this study is to develop automated machine learning (ML) classifiers that can identify subtle differences in affective states between young children who do stutter (CWS) and who do not stutter (CWNS) during a stressful scripted speech and a narration task.}
\par
%Anything in our environment can trigger an emotion (an affective state), which can be defined as physiological and psychological change in response to a stimulus \cite{liu2017many,russell2003core}. 
Emotions (i.e., affective state) are temporary, last for a short time, and are complex psychophysiological constructs composed of two underlying dimensions: valence and arousal \cite{kuhbandner2011dissociable,lang1993looking}. Valence is defined as the positive to negative evaluation of the subjectively experienced state \cite{harmon2011expression}. Arousal measures the intensity of the affective state ranging from calm to highly excited or alert \cite{lang1995emotion,bradley2007emotion}. 
\par
%Physiological fingerprints of affective states have been studied extensively \cite{liu2017many}. 
Psychological arousal of an individual is observed as spontaneous responses in physiology due to an external (seeing a scary picture) or an internal (one’s own thought) stressor \cite{liu2017many}. These responses are spontaneous and can manifest themselves as changes in heart rate \cite{magagnin2010heart}, electrodermal activity (skin conductance) \cite{mauri2010psychophysiological}, etc. The autonomic nervous system (ANS) is responsible for directing these physiological responses. The sympathetic nervous system (SNS), one branch of the ANS, directs the ‘fight or flight’ response. It stimulates the body to respond to a stressful situation by the elevation of physiological parameters like heart rate, respiration rate, blood glucose levels. The other branch of the ANS, the parasympathetic nervous system (PNS), directs the ‘rest and digest’ response. It conserves the body’s natural activity and relaxes the individual once a stressful situation has passed. The PNS leads to a decreased arousal by reducing the heart rate and respiration rate.
\par
\begin{figure}[]

         \centering
         \includegraphics[width=\textwidth]{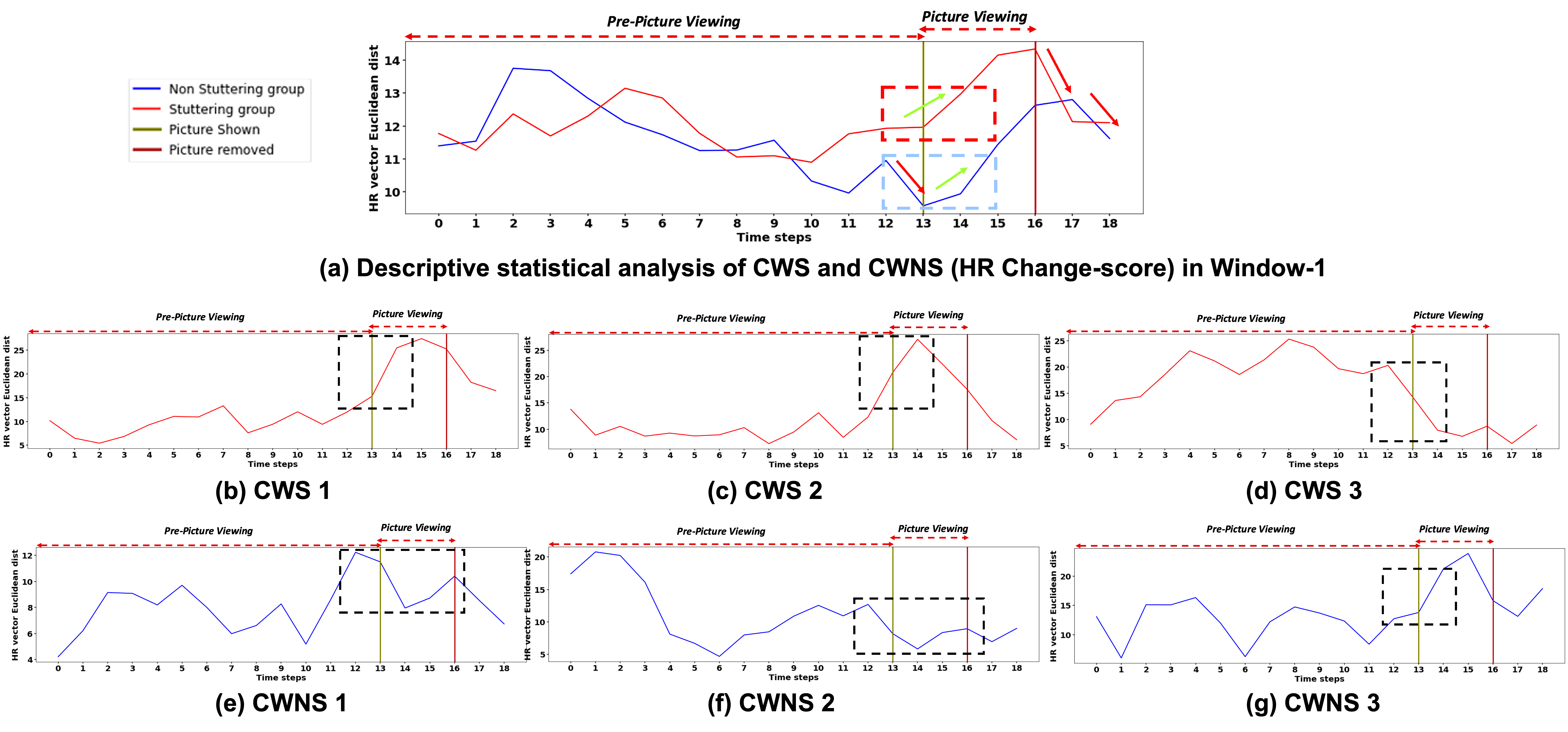}
         \caption{(a) Shows the descriptive statistics of HR change-score Euclidean-distance-feature for CWS and CWNS in window 1 of the scripted dataset (section \ref{Change-score-features}), which captures the participants’ physiological response while waiting for a picture, watching the negatively-valenced picture, preparing to speak in stressful conditions. (b), (c), and (d) are three 20s physiological response (HR change-score Euclidean-distance-feature) examples from three different CWS participants. Similarly, (e), (f), and (g) are three examples from three different CWNS participants. In all figures, the dotted rectangles mark the CWS or CWNS indicative distinctive patterns present in the respective 20sec window.}

     \label{HR_descriptive-stat}
     \vskip -3ex
\end{figure}
The two branches of the ANS interact to coordinate our physiological responses. The interaction of the SNS and PNS, the two branches of the ANS, is demonstrated in children's cardiovascular response to a stressful situation shown in Figure \ref{HR_descriptive-stat}(a). It is a descriptive statistical comparison of `mean heart rate (HR) change-score Euclidean-distance-feature' of the CWS and CWNS participants before, during, and after seeing a negatively-valenced picture. For CWS participants, while seeing the picture, SNS causes the mean HR to increase. That is marked by a green arrow. Once the stressful situation is over (i.e., removal of the picture), the PNS reduces the mean HR back to the resting state. That is marked by a red arrow. However, the mean response of the CWNS population follows a different trend. While exposed to an external stressor (i.e., negatively valenced picture viewing), CWNS's PNS reduces the mean HR even lower (marked by a red arrow), and subsequently, the SNS causes the mean HR to increase (marked by a green arrow). Such a response is called \textit{freezing response} \cite{alm2004stuttering}, which is a paradoxical decrease in HR during stressful situations. Eventually, when the stressful situation is over (i.e., removal of the picture), the PNS reduces the mean HR back to the resting state (marked by a red arrow). Notably, not all participant's physiological response follow their corresponding group's (CWS or CWNS) trend. For example, in figure \ref{HR_descriptive-stat}(d), the `CWS 3' participant's HR reduces while seeing the negatively-valenced picture, and in figure \ref{HR_descriptive-stat}(g), the `CWNS 3' participant does not experience freezing response, rather HR increases while seeing the picture; these responses are different compared to the mean response from the CWS and CWNS population, respectively. Traditional statistical methods describe the general trend of various physiological parameters as they operate on averages (e.g., calculation of the mean across the specific condition). However, they fail to identify the personalized and group-wise fine-grain, second-by-second differences (i.e., distinctive patterns) in physiological parameters between CWS and CWNS. Our observation and discussion above motivate the need for such analysis that the presented study performs.
%Such observations motivate the use of machine learning (ML) classification approaches that offer a new way to examine physiological response data which includes fine-grain, second-by-second and personalized assessment of physiological parameters. %\textcolor{red}{This level of analysis could inform our understanding of how emotional arousal could contribute to the development of stuttering overall and whether subtle differences in physiological arousal can lead to immediate changes in speech rhythm and fluency in the form of sound repetitions, prolongations or silent tense pauses, the core features of stuttering.}
\par
Speech production is a complex process which requires precise coordination vocal tract while simultaneously processing cognitive-linguistic information. Social engagement, including regulating own emotions and responding appropriately to one’s communicative partner, is also inherent to spoken communication. Naturally, speech production can be affected by the speaker’s physiological arousal.  Studies \cite{arnold2014autonomic,kleinow2000influences,choi2016emotional,tumanova2019autonomic,zengin2015sympathetic,zengin2018sympathetic,jones2014autonomic} have shown that young children who stutter are especially vulnerable to such influences.
\par
Moreover, stuttering is a neurodevelopmental speech disorder \cite{smith2017stuttering} that emerges in early childhood (between the ages of 2 and 4), hence, it is essential to examine the effects of physiological arousal on speech characteristics in young children as opposed to adults. Given that preschool-age is the time when essential communication skills are undergoing most significant development and also when some children develop stuttering, it is essential for our understanding of stuttering to examine young children’s physiological response during speech production. 
%Young children (whose speech motor control is not yet mature) and those speakers who have speech disorders may be especially vulnerable to these influences. 
\par 
\textit{This study presents an interpretable AI approach to identify the second by second fluctuations and pattern differences in physiological arousal of preschool-age children who stutter compared to others who don't during various speaking tasks.} 
This level of analysis could inform our understanding of how emotional arousal could contribute to the development of stuttering overall and whether subtle differences in physiological arousal can lead to immediate changes in speech fluency/disfluency and speech articulation.
\par 
Moreover, the developed machine learning classifiers identify personalized distinctive situational physiological arousal for each 20sec physiological sensing data. Hence, they can be leveraged for remote, continuous, and real-time assessment of CWS’s physiological arousal and may lead to automated, personalized, and just-in-time interventions to mitigate their physiological arousal, consequently mitigating stuttering disfluency.
\par 
\textit{What follows is an overview of the nature of stuttering, the autonomic nervous system activity in response to speaking in CWS and the potential role of physiological reactivity in speaking and the development of stuttering (section \ref{arousalinchildren-stat}). Further, we discuss the deep learning frameworks to understanding of children’s physiological response during speech (sections \ref{priorwork-ML}, and \ref{prior-deepML-stutter}). We end the introduction with the study challenges (section \ref{challenges}), and our research questions, problem statement and contributions (section \ref{problem_statement}).}

\subsection{Physiological Arousal in Children who Stutter}\label{arousalinchildren-stat}
Stuttering is believed to be a multifactorial condition where multiple factors interact and contribute to stuttering onset in early childhood and its later development to its chronic form \cite{smith2017stuttering}.  Research shows that speech production leads to increased autonomic arousal in both adults \cite{het2009neuroendocrine,kirschbaum1993trier,weber1990autonomic}, and children \cite{arnold2014autonomic,kleinow2000influences,choi2016emotional,tumanova2019autonomic,zengin2015sympathetic,zengin2018sympathetic,jones2014autonomic}. In this study we were interested in examining whether speaking tasks that vary in linguistic complexity are inherently more stressful and associated with higher physiologic arousal for preschool-age children who stutter. We were also interested in examining any patterns in physiological reactivity that distinguish children who stutter from their typically fluent peers. 
Psychophysiological research to date offered mixed findings regarding whether preschool-age children who stutter differ in their autonomic arousal during speech production from their peers who do not stutter.
\par
Literature from speech science generally indicates that preschool-age children who stutter do not have an elevated autonomic arousal during such speaking tasks as picture naming, picture description, and non-word repetition \cite{arnold2014autonomic,kleinow2000influences,choi2016emotional,tumanova2019autonomic,zengin2015sympathetic,zengin2018sympathetic,jones2014autonomic,jones2017executive,walsh2019sympathetic,walsh2019physiological}. However, differences in autonomic arousal between children who do and do not stutter based on the children’s age \cite{zengin2015sympathetic} and complexity of the speaking task \cite{tumanova2019autonomic} have been reported. Further, among children who stutter, differences related to children’s stuttering chronicity\cite{zengin2018sympathetic}, and speech fluency \cite{walsh2019physiological} have also been observed. 
Importantly, studies published to date relied on traditional statistical approaches in examining potential differences in physiological arousal between children who do and do not stutter. 
However, identifying fine-grain, second-by-second, and personalized differences in physiological parameters between CWS and CWNS groups during speaking tasks has yet to be addressed.

\subsection{Deep Learning in Physiological Arousal and Stress Detection}\label{priorwork-ML}
Various ML approaches (e.g., SVM, CNN, RNN, RCNN) have been developed to detect arousal/stress through physiological sensors \cite{albertetti2021stress,hssayeni2021multi,maier2019deepflow} from non-stuttering individuals. For example, a study \cite{rastgoo2019automatic} evaluated a convolutional neural network (CNN) and long short-term memory (LSTM), taking a combination of ECG signal features, vehicle, and contextual data as input to predict driver stress with an accuracy of 92.8\%. Another study \cite{albertetti2021stress,maier2019deepflow} used CNNs and Recurrent CNNs to detect the stress using physiological parameters such as electrodermal activity (EDA) as features and achieved an accuracy of 67.50\% and F1-score of 0.71 respectively in detecting stress. Table \ref{prior_work} in Appendix \ref{related-work-appendix-stress} summarizes the recent literature on physiological sensing-based stress assessment.

\subsection{Deep Learning and Physiological Sensing for Stuttering Speech Disfluency detection}\label{prior-deepML-stutter}
To the best of our knowledge, no study has developed ML classification approaches to differentiate the physiological responses between CWS vs. CWNS during perceptually fluent speaking tasks like scripted phrase repetition under an external stressor and spontaneous narration conditions. However, a study \cite{villegas2019novel} on adults with stuttering (AWS) has developed a multi-layered perceptron (MLP) neural network-based disfluency classification system. It used respiration signals as input and achieved an 82.6\% accuracy in differentiating the physiological response of `AWS during disfluent speech production’ vs. `AWNS during fluent speech production.’ In this study \cite{villegas2019novel}, the participants were asked to read aloud a 677-syllable text extracted from a Spanish story. The speech-language pathologists observed the experiments and provided fine-grain annotations of stuttering disfluency events of the AWS, such as sound/syllable repetitions, sound prolongations or silent tense pauses. 
%Even though fine-grain annotations are needed to effectively train supervised learning classifiers (e.g., DNN, CNN, LSTM), they are time-consuming, costly, and dependent on the annotators’ expertise. 
\textit{Notably, in contrast, in our data (section \ref{study_design}), children are talking perceptually fluently for approximately 97\% of the duration. Hence, the presented study focuses on differentiating the physiology of the CWS vs. CWNS during both perceptually fluent and disfluent speech production.} Though statistical-based studies (discussed in section \ref{arousalinchildren-stat}) have shown that there are group-wise differences in the physiological response of CWS vs. CWNS during perceptually fluent speech production, no study to our knowledge has examined or identified the second by second salient and distinguishable patterns present in CWSs’ vs. CWNSs’ physiology during their fluent speech. Since such patterns are unknown, it is not possible to annotate our data at the fine-grain level, which makes our classification a challenging task \cite{happy2019weakly,wu2015deep}.  

%This paper adopted a weakly supervised learning approach that does not require expert annotations to address the challenge.

\subsection{Study Challenges}\label{challenges}
This paper aims to develop automated ML classifiers that can identify the subtle differences in affective states between CWS and CWNS during the stressful scripted speech and narration tasks. This section discusses the challenges of developing such a classifier effectively.
\par
\subsubsection{Weakly Labeled Data:} A challenge is, though prior statistical studies \cite{zengin2015sympathetic,tumanova2019autonomic} established the existence of physiological parameter differences in CWS and CWNS, none of them identified the CWSs’ distinctive physiological signal patterns.
Moreover, preschool-age children cannot self-assess and report on their physiological state due to their young age \cite{henderson2015behavioral}. Also, changes in arousal would not always be accompanied by observable changes in behavior.
%Moreover, self-labeling from preschool-age participants is unreliable, and their temperaments and behavioral inhibitions make data labeling by external observers difficult \cite{henderson2015behavioral}.
Ergo, our data is `weakly labeled’, meaning we do not have the precise labels of CWS's distinctive physiological signal patterns. For example, figure \ref{HR_descriptive-stat}(b)-(g) show six children’s 20s cardiovascular response (i.e., HR change-score) window. The dotted rectangles show the children’s class (CWS or CWNS) indicative distinctive patterns. These patterns are $2-5$s in duration and may appear in any timestamps of the 20s window. Our data annotations only provide information about the data belonging to a CWS or CWNS individual; the distinctive patterns and their timestamps in the detection window are unknown. Additionally, our datasets are limited in size. There were 180 and 200 picture-viewing events for CWS and CWNS during the scripted speech experiment. Supervised learning classifiers fail to learn the above discussed subtle, sparse and independent physiological sensing patterns from `weakly' labeled data, specifically while limited in size.
%Notably, physiological sensing patterns that differentiate CWS and CWNS can comprise a small segment (i.e., fraction) of the total detection window and are independent of other sensory stream segments. An example with the EDA signal is shown in the figure \ref{EDA_variations}. Figures \ref{EDA_CWS} and \ref{EDA_CWNS} show the EDA standard deviation (EDA-std) data during the negatively-valenced picture viewing experiment (discussed in section \ref{procedure}) for stuttering and non-stuttering participants. It can be observed that the affective stimuli (negatively-valenced image viewing) at 13s timestamps caused a stiffer EDA-std increase (marked with red arrows with label ‘A’) in the stuttering participant than the other. The differentiating segments are 2-2.5s long in duration. The rest of the EDA-std segments are arbitrary and do not convey any differentiating patterns.
\par
\subsubsection{Modality-wise Distinctive Patterns:} 
\begin{figure}
     \centering
     \begin{subfigure}[t]{0.49\textwidth}
         \centering
         \includegraphics[height=1.5in,width=\textwidth]{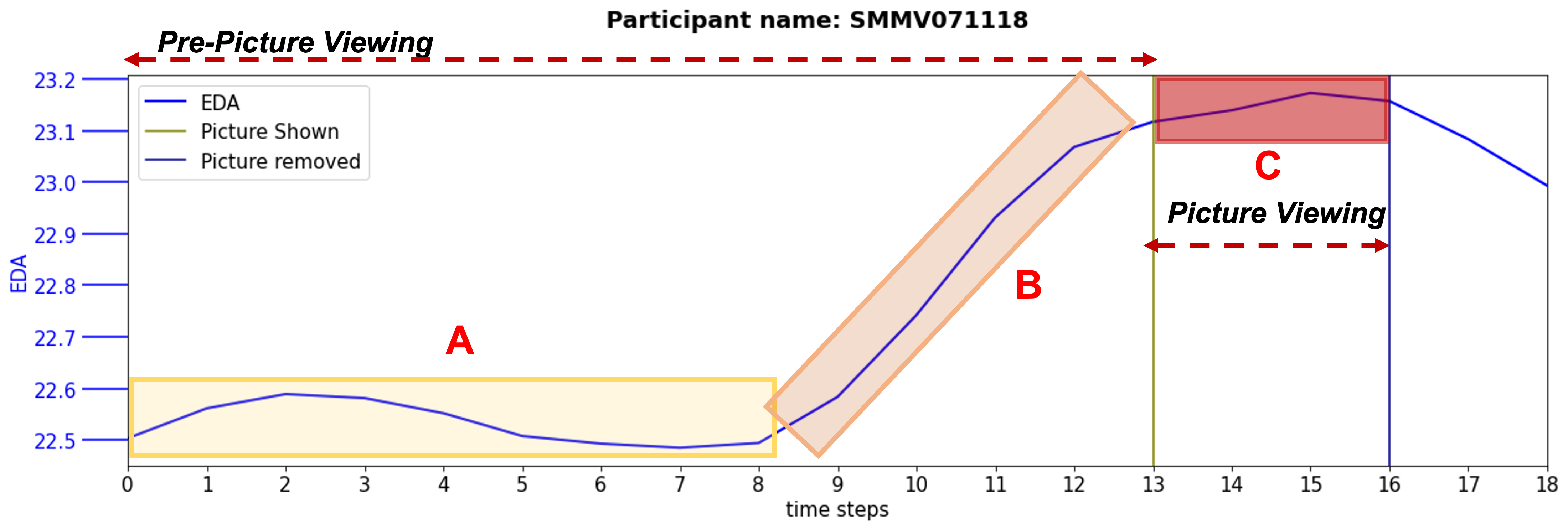}
         \caption{EDA variations}
         \label{EDA_variation_CWS}
     \end{subfigure}
     \begin{subfigure}[t]{0.49\textwidth}
         \centering
         \includegraphics[height=1.5in,width=\textwidth]{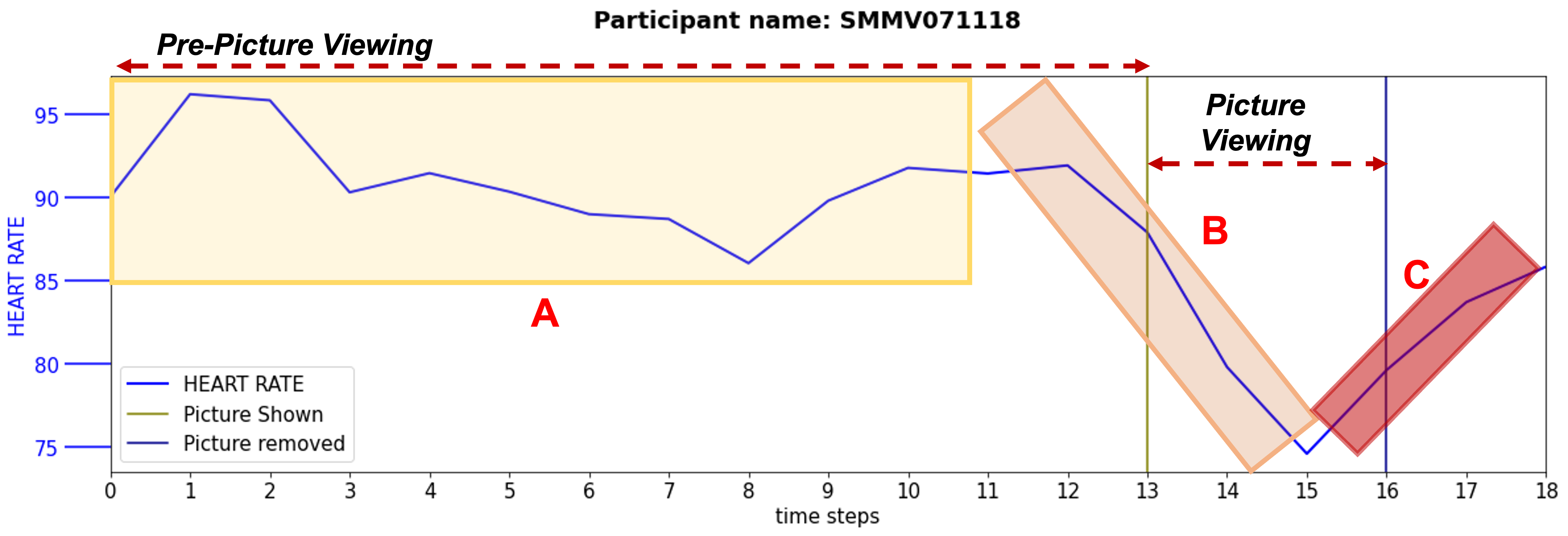}
         \caption{Heart rate variations}
         \label{HR_variation_CWS}
     \end{subfigure}
     \caption{EDA and HR signals during negatively valenced picture viewing for a CWS participant. The `Yellow’ regions (i.e., region A) of the signals do not convey any arousal indicative pattern. `Orange’ regions (i.e., region B) show the signal patterns indicative of low-to-high arousal transition, and the `red’ regions (i.e., regions C) show the high arousal response. }
     \label{variations}
\end{figure}

This paper evaluates multiple physiological modalities (e.g., HR, EDA and Respiration activities) to measure the participants’ (i.e., CWSs’ and CWNSs’) physiological response. A notable observation is that arousal indicative sparse patterns do not simultaneously emerge in each modality. Figure \ref{variations} shows the EDA and HR signals of a CWS participant during a negatively valenced picture viewing. As shown in Figure \ref{EDA_variation_CWS}, a low-to-high arousal transition (i.e., region B) appears in the EDA during 8s to 13s timestamps while the participant is waiting for a picture to be viewed (anticipation effect \cite{greenberg2015anticipation}), and the participant is experiencing high arousal during the negatively valenced picture viewing (i.e., region C). However, according to Figure \ref{HR_variation_CWS}, a low-to-high arousal transition (i.e., region B) appears in the HR during picture viewing (12s-15s timestamps), where the HR decreases rapidly (\textit{freezing response} \cite{alm2004stuttering}). High arousal (i.e., region C) in HR is depicted as the subsequent increase in HR after the freezing response. 
Notably, our `weakly’ labeled data do not have the fine-grain annotations of such arousal indicative of sparse patterns in any of the modalities. The above-discussed observation motivates that \textit{our developed classifier needs to extract CWS's distinctive sparse patterns independently from each modality, without any available annotations (for such patterns) during training.}
\par 
\subsubsection{Capturing Cross-modality Dependency: } Previous studies \cite{blain2010cardiorespiratory,kettunen1998synchronization} have shown that the correlations between physiological parameters such as HR, EDA and Respiration activities are effective attributes in stress \cite{albertetti2021stress,hssayeni2021multi,maier2019deepflow} or emotional valence \cite{shu2018review} detection. Moreover, the HR-EDA synchronization is positively associated with variability of arousal responses \cite{kettunen1998synchronization}. Hence, our solution approach must capture the cross-modality relationships as well. 
\par
\paragraph{In conclusion,} our solution needs to identify CWS's distinctive sparse patterns independently from each physiological modality without any available annotations of such patterns, and capture and leverage the cross-relationships of the identified modality-specific sparse patterns for effective CWS vs. CWNS classification.

\subsection{Problem Statement and Contributions}
\label{problem_statement}
%As discussed above, the differentiating patterns of CWS vs. CWNS can be small, independent, and sparse, and the data is `weakly’ labeled.This paper adopted a weakly supervised learning approach that does not require expert annotations to address the challenge.
In particular, the study aims at investigating the following key questions. To our knowledge, no previous study has investigated these questions through machine learning approaches. 
\begin{enumerate}

    \item \textbf{Do the CWS and CWNS show different physiological responses to external stressors?} \label{q1}\\
    This question aims to find differences in the physiology of the CWS and CWNS group under the arousal inducing condition when the participants are viewing negatively-valenced pictures. To investigate, we developed and evaluated classifiers to differentiate the CWS vs. CWNS from the 20s ‘window-1’ signals of the scripted dataset (discussed in sections \ref{Res-Q1-Eval} \& \ref{Preprocessing:event-window-extraction}).
    
    \item \textbf{Do the CWS and CWNS show different physiological responses while perceptually fluently talking under stressful conditions?}\label{q2}\\
     %To our knowledge, no previous study has investigated this question. 
     This research question aims to understand the differences (i.e., if any) in the physiology of the CWS vs. CWNS groups while talking under stressful conditions (i.e., after viewing negatively-valenced pictures). To investigate, we developed and evaluated classifiers to differentiate the CWS vs. CWNS from the 20s ‘window-2’ signals of the scripted dataset (discussed in sections \ref{Res-Q2-Eval} \& \ref{Preprocessing:event-window-extraction}).
     
    \item \textbf{Do the CWS and CWNS show differences in physiology under rest or baseline condition?}\label{q3}\\
    To investigate, we developed and evaluated classifiers to differentiate the CWS vs. CWNS from the 20s baseline/neutral condition physiological signals of the scripted dataset. (discussed in section \ref{Q3:Evaluationanalysis})
    
    \item \textbf{Do the CWS and CWNS show different physiological responses during spontaneous narration?} \label{q4}\\
    The narration task is linguistically and cognitively demanding since the children develop new context or storylines and articulate them in speech. Therefore, this research question investigates whether the CWS show different physiological responses than the CWNS while performing such narration task. To investigate, we developed and evaluated classifiers to differentiate the CWS vs. CWNS from the 20s windows of narration task condition of the free-speech dataset (discussed in section \ref{Res-Q5-Eval}).
    
\end{enumerate}

\subsubsection{Contribution in Classification:}
To investigate the above-mentioned questions, we develop and evaluate a \textit{novel Modality Invariant-MIL (MI-MIL) (section \ref{MI-MIL}) classifier}.
\begin{enumerate}

    \item To address the weakly labeled data challenge and identify modality-wise distinct patterns (section \ref {challenges}), \emph{MI-MIL applies modality-wise multiple-instance-learning (MIL) paradigm in each physiological modality independently.} MIL paradigm is designed to extract sparse and subtle patterns from weakly labeled data (i.e., without any fine-grain annotations of the region, timestamps, or duration of the patterns in the data).
    
    \item To capture the cross-modality-relations (section \ref {challenges}), \emph{MI-MIL presents a novel modality-fusion network that identifies the cross-relations of each modality’s CWS indicative sparse patterns.}

    \item Our evaluations discussed in section \ref{Classifier-Development-EVAL} show, the presented approach outperforms the supervised learning classifiers, and the recent state-of-the-art MIL approach (attention-based MIL) significantly.
    
    \item Our evaluation demonstrates, MI-MIL is real-time executable in scalable resource constraint devices: NVIDIA Jetson Nano and Google pixel 6 smartphone.
    
\end{enumerate}
Developed MI-MIL models’ high efficacy in addressing each of the four questions ($q1$-$q4$) indicates the existence of physiological signal patterns that differentiate the binary categories, and our classifier can identify them. 
\par 
\subsubsection{Contributions in Dataset Collection and Physiological Features:}

\begin{enumerate}

\item We collected \textit{two datasets (section \ref{study_design}) containing CWS and CWNS’s physiological responses} (i.e., physiological sensing parameters: electrodermal activity - EDA, heart rate - HR, respiratory rate - RSP-rate, and respiratory amplitude - RSP-amp) while speaking in two challenging conditions: speaking under stressful conditions (after experiencing external stressors) and a linguistically and cognitively demanding narration task where the children need to spontaneously develop new context and concepts. 

%A challenge utilizing raw physiological sensing parameters is, they are subjective, and different individuals' dissimilar affective states can have similar raw physiological parameters. 
\item Motivated by the state-of-the-art behavioral science studies, we extracted \textit{a novel vector-distance-based representation of change-scores features} of physiological parameters that captures the fluctuation of one’s physiological response in a target condition than their neutral condition (section \ref{Change-score-features}). Presented change-score representation significantly outperforms the representation used in literature (Appendix \ref{change-score-eval-sec}).

\item We evaluated ML models for both raw and change-score physiological features. They give us insights into the CWS vs. CWNSs’ physiological-patterns and fluctuations-differences. Furthermore, we evaluated and ranked the raw features according to their discriminative capabilities (section \ref{eval-approach-components}). The feature importance ranking is in line with our observation of Shapley visualization (section \ref{interpretability}).
\end{enumerate}
\par 
\subsubsection{Classification Interpretation:} This study is the first-of-its-kind to analyze, interpret, visualize and discuss the fine-grain, second by second, temporal, and distinctive physiological response patterns of CWS from CWNS during speech production in different challenging conditions. The developed MI-MIL classifiers are black-box models; hence we employ SHAP \cite{chen2019explaining} ML model explainers (detailed discussion is in section \ref{interpretability}) to extract and visualize the \emph{group-wise} and \emph{personalized} situational physiological arousal patterns.
Identifying and visualizing group-wise patterns would enhance our understanding of stuttering etiology and development. Personalized pattern identification would enable remote, continuous, and real-time assessment of stuttering children's physiological arousal, which may be used clinically to develop personalized emotion regulation strategies (e.g., biofeedback, mindfulness intervention), resulting in an improvement in speech fluency.

\section{Description of the Dataset and our data Collection Procedure}
\label{study_design}
We collected two datasets comprising preschool-age children's (both CWS and CWNS) physiological response (i.e., physiological sensing parameters) while performing two different speaking tasks. 
%two research datasets collected from preschool-age children for two separate projects conducted by the X Stuttering Research Laboratory (e.g., \cite{tumanova2019autonomic,tumanova2020role}). 
Due to the nature of speaking tasks, we refer to the first dataset as “free speech dataset” and to the second dataset as “scripted dataset” consistently throughout the manuscript. Study participants and procedures are explained below.
\par
\textit{Participants:}
Participants in both datasets were preschool-age children (mean age: 50.3 months, std: 9.14). The study procedures were approved by the Syracuse University Institutional Review Board. All data collection procedures took place in our Laboratory over two visits. During the first visit participants were administered standardized tests of speech and language to ensure age-appropriate speech articulation and language scores, and passed a pure-tone hearing screening. Participants’ speech fluency was assessed by a licensed speech-language pathologist and the diagnosis of stuttering was established using evidence-based diagnostic criteria \cite{tumanova2014speech,yaruss1998evaluating}. All psychophysiological data were collected during the second visit.  
The free-speech dataset comprised data from 35 preschool-age children (age range: 36-67 months). Among the participants, 16 were CWS, (13 boys and 3 girls), and 19 were CWNS (12 boys and 7 girls).
The scripted dataset comprised data from 38 preschool-age children (age range: 38-69 months). Among the participants, 18 were CWS, (16 boys and 2 girls) and 20 were CWNS (16 boys and 4 girls).

\subsection{Data-Collection Experimental Procedure}
\label{procedure}
We explain the common procedures followed in data collection for both datasets first, then we describe the dataset specific differences in procedures. Upon arrival at the lab, participants played and spoke with the examiner for about \textit{15 minutes} to get them acquainted with the lab. Then, they were seated in a child-sized chair, in front of a computer screen. Hypoallergenic electrodes were attached to the skin at the suprasternal notch of the rib cage and at the 12th rib laterally to the left for acquisition of the electrocardiogram \cite{venables1980electrodermal}. A strain gauge transducer designed to measure respiratory-induced changes in thoracic or abdominal circumference (model TSD201, Biopac Systems, Inc.) was used to record respiratory effort. The transducer was positioned around the participants’s chest for acquisition of the respiration waveform. The electrodermal activity was recorded with electrodermal response transducers (model TSD 203, Biopac Systems, Inc.) which included a set of two Ag-AgCl electrodes with incorporated molded housings designed for finger attachment. The response transducers were filled with an isotonic electrolyte gel and were placed on the volar surfaces of the middle phalanges of the two fingers of the participants’ right hand. 
After the sensors were placed, the participants’ baseline psychophysiological data were collected first followed by the experimental conditions. The conditions are explained below. 

\subsubsection{Baseline Condition:}
\label{baseline_condition}
For both datasets, to establish a pre-experimental baseline for each participant’s resting skin conductance level, breathing rate and heart rate, participants viewed an animated screensaver of a three-dimensional fish tank for \textit{four minutes}. This procedure has been successfully implemented in prior studies to establish baseline psychophysiological levels in preschool-age children \cite{jones2014autonomic,tumanova2019autonomic}.
\subsubsection{Experimental Condition - Free Speech Dataset:}
The experimental condition in this dataset was a picture description task, which lasted about \textit{10 minutes}. During the picture description task, participants were shown pictures from a wordless storybook about a boy, a dog, and a frog by the author \textit{Mercer Mayer, Frog Goes to Dinner} \cite{mayer1974frog}. To keep the narrative elicitation procedure consistent between the participants, the examiner was not allowed to ask specific questions about the picture but could only prompt the participant to tell them what was happening in the picture by saying “Let’s look at this picture. Tell me what is happening here.” The examiner was instructed to provide no more than three such elicitation prompts per picture. Participants who stuttered did show some (3\% of the total speech) stuttering events, such as sound repetitions and prolongations, during this experiment.
\subsubsection{Experimental Condition - Scripted Dataset:}
The experimental condition in this dataset also lasted approximately \textit{10 minutes} and involved negatively-valenced picture viewing and phrase repetition. Specifically, the participants were shown $10$ negatively-valenced pictures from the International Affective Picture System\cite{lang1997international} and were asked to repeat a target phrase “Buy Bobby a puppy” (BBAP phrase) after a pre-recorded prompt presented over the speakers. Picture presentations were interspersed with speaking such that per each picture shown the participants were asked to repeat the target phrase $3$ times. Figure \ref{window} shows the chronological order of events for one negatively-valenced picture viewing in the scripted dataset. 
None of the participants showed any stuttering disfluency events during this experiment.
\subsubsection{Data Acquisition: Preprocessing and Cleaning:}
\label{data-acquistion}
The respiratory, electrodermal, and cardiac activity were acquired simultaneously using the Biopac MP150 hardware system (Biopac Systems, Inc.) and recorded using Acqknowledge software (ver. 4.3 for PC, Biopac). Respiratory effort (RSP), electrodermal activity (EDA), and electrocardiogram (ECG) signals were sampled at 1250 Hz during the baseline and experimental conditions. The butterworth high pass filter\cite{rehman2013survey} was applied to the raw signals to remove the noise and baseline drift.

\subsection{Extraction of Event Detection Windows:}\label{Preprocessing:event-window-extraction}
This paper detects the affective state differences from 20s windows; since in the scripted dataset, children took on avg. 20s to utter BBAP phrase $3$ times following a picture viewing. For the \textit{free-speech dataset}, baseline-level data collection and picture description sessions lasted approximately 4min and 10min. We segment the sessions into 20s windows with a5s overlap (15s hop-length). Physiological sensory streams from each 20s window from baseline and picture description session represent the participant’s physiological response in the neutral and narration (i.e., linguistic and cognitively demanding) conditions.
Similarly, in the \textit{scripted dataset}, baseline-level data collection lasted approximately 4min, which was segmented into 20s windows. Sensory signals from these 20s windows represent participants’ physiological response in the neutral condition. During the speaking task, for each negatively-valenced picture viewing, we extracted two 20s windows: (1) during picture viewing; and (2) after the picture viewing while the participants were repeatedly saying the scripted phrase (“Buy Bobby a puppy”).
\begin{figure}[ ]
  \centering
  \includegraphics[width=0.85\textwidth]{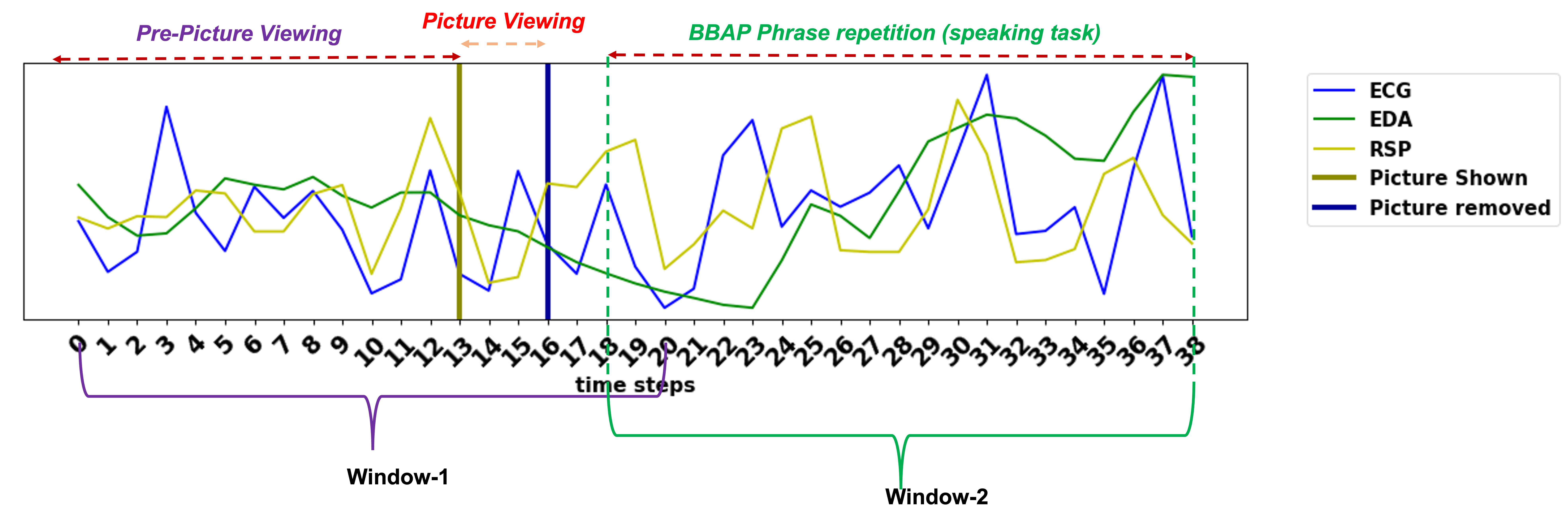}
  \vskip -3ex
  \caption{Chronological order of events for one negatively-valenced picture viewing in the scripted dataset.}
    \label{window}
    \vskip -3ex
\end{figure}
As shown in figure \ref{window}, the picture was flashed on the screen for 3 seconds (i.e., red region), after which the computer prompts the sentence "Buy Bobby a puppy" (BBAP). After that, participants started to speak, which took approximately 20s. The first window starts  at time step 0 (where in each time step is a 2s window)  and the image is shown at time step 13. Hence, window 1 of the scripted dataset captures the participants’ physiological response while waiting for a picture, watching the pictures, preparing speech in stressful conditions. That is, it captures the participants’ progression from normal to stress state.
Window 2 started at time-step 18 and ended at time step 38. Hence, Window 2 of the scripted dataset captures the participants’ physiological responses while speaking in stressful conditions. (It is important to note that window-1 and window-2 both consist of nineteen time-steps or 2s instance windows each have 1s overlap resulting in 19 time-steps for a 20s window).
The data collection sessions were segmented into 20s windows with a 5s overlap; hence for the free speech dataset, we had a total of 1510 windows. For the scripted dataset, there were 752 windows in total; wherein there were 376 window-1 and window-2 each.

\subsection{Physiological Features Extraction}
\label{feature_extraction}
As discussed in the previous section each of the 20s signals was divided into nineteen segments with 2s duration and 1s overlap, and features are extracted from the 2s segments. We evaluated the event detection models using two categories of features: (a) raw features and (b) change-score features.

\subsubsection{Raw Physiological Features}
\label{Raw_physio_features}
Psycho-physiological features relevant to affective states and stuttering individual’s physiological responses are extracted from each 2s segments. We extracted the \textit{heart rate (HR)} from the raw ECG signal, since a recent study \cite{tumanova2020emotional} of heart rate (HR) in relation to stressful situations indicates that children who stutter showed a significantly higher HR than CWNS. \textit{Electrodermal activity (EDA)} signals increases and shows spontaneous fluctuations during arousal \cite{bach2011dynamic,nikolic2011measuring}. Hence, filtered EDA is one of the extracted features. Moreover, studies have shown respiratory distributions vary in individuals with stuttering vs. non-stuttering and different speech-fluency levels \cite{tasko2007speech}. To capture the children’s respiratory patterns, we extract the \textit{respiratory rate (RSP-rate) and respiratory amplitude (RSP-amp)}\cite{schneider2003respiration} as raw physiological features from the raw RSP signal.
\par
Following previous studies \cite{salekin2017distant,siirtola2019continuous}, two-level physiological features (Low-level and high-level Descriptors) are extracted, allowing the ML models to capture signal characteristics in different granularity levels. Four low-level descriptors (LLD) features: heart rate (HR), electrodermal activity (EDA), respiratory rate (RSP-rate), and respiratory amplitude (RSP-amp), were extracted at 0.8-millisecond intervals from each 2s segment. Six high-level descriptors (HLD) functionals: min, max, std, var, mean, and median, are applied to the LLDs to extract the feature representation for each of the four LLDs, totaling <24> HLD features, are extracted from each 2s segment. In total, we extract $19\times24$ raw physiological features from the 20s event detection window.

\subsubsection{Change-score Features}\label{Change-score-features}
State-of-the-art behavioral science studies evaluate change-score \cite{chiu2015complexity,clifton2019correlation,jones2017executive} features to understand the psychophysiological changes on individuals in response to different affective states (e.g., arousal). 
A change score is the difference between the value of a variable/feature measured at one point in time ($Y_t$) from the average value of the variable for the same unit at the baseline-level condition ($Y_b$). $Y_t$ is called the `post-score,’ $Y_b$ is the `baseline-score,’ and the difference between
$Y_t$ \& $Y_b$ is the `change-score’. This study extracts change-scores of HR, EDA, RSP-amp, and RSP-rate LLD-features from each 2s segment. The post-, baseline-, and change-scores of these physiological features are represented as vectors.
\par 
\paragraph{Post-scores} are calculated from each 2s segments in different non-baseline scenarios. Each of the LLD features' is represented as a 6-dimensional vector (i.e., one dimension for each HLD-functional) quantifying an individual's physiological response. These 6-dimensional vectors %$\vec{Y_{HR}},\vec{Y_{EDA}},\vec{Y_{RSP-amp}},\vec{Y_{RSP-rate}}$ 
are the respective LLD features' post-scores.
\par 
\paragraph{Baseline-scores} are calculated from all of the 2s segments in the individual's baseline condition. For each LLD feature (HR, EDA, RSP-amp, and RSP-rate), we consider the mean of its 6-dimensional HLD vectors from all baseline-condition 2s segments as its `baseline-score' vector. Meaning, we extract four 6-dimensional baseline-scores/vectors for each individual, representing the average LLD-features values in their neural condition.
\par 
\paragraph{Change-scores} are the vector differences between the post-scores and baseline-score. For each 2s non-baseline segment, it quantifies the difference in an individual's physiological response regarding their baseline (i.e., neutral) response.  
%Different individuals may have different physiological responses in the baseline-level condition. For example, an individual may have above-average EDA in the baseline-level condition. However, a machine learning classifier trained on raw physiological features (section \ref{Raw_physio_features}) would be unaware of such exception and infer the individual's neutral state as stress-state. The change-score features eliminate such biases.
\par 
In this study, the vector difference between the post-score and baseline-score is measured by two matrices: cosine similarity and the euclidean distance. They are the two most common matrices to measure vector difference used in machine learning \cite{vijaymeena2016survey,kryszkiewicz2014cosine,gomez2019self,wang2005euclidean}.
\par 
For each of the four LLD features HR, EDA, RSP-amp, RSP-rate, we calculate two change-score values (euclidean distance and cosine similarity), totaling eight change-score features extracted from each 2s small-signal segments. In total, we extract $19\times8$ change-score features from the 20s event detection window.

\subsubsection{Comparison of Features:} \label{Feature-extraction-comparison} Raw physiological features (HLDs) \textit{captures the signal amplitudes and patterns of individuals in different conditions}. Hence, affective state detection classifiers learn the physiological parameter values and time progression relevant to different target classes (e.g., arousal). However, a limitation is that individuals' physiological responses in the neutral condition can be dissimilar. Hence, a classifier trained on raw physiological signals may misclassify an individual’s neutral state to aroused state if their baseline-level physiological signal attributes are different from the average population. Change-score features eliminate such bias since they \textit{capture the difference in an individual's physiological response in different conditions compared to their neutral state.} However, in doing so, change-score features lose fine-grain information of the signals (e.g., signal amplitudes, std, etc.). Hence, \textit{this paper trained and interpreted classifiers using both raw and change-score physiological features to understand how their `attributes (amplitudes and patterns)' and `fluctuations compared to the neutral condition'} indicate children’s mental states during phase repetition under aroused condition and narration tasks.

\section{Methodology: CWS vs. CWNS Psychophysiological Arousal Detection}
\label{Methodology}
\subsection{Approach Design Choices}\label{Method-challenges}
%make this section shorter. Refer 4.2 sections.
This section discusses how our presented MI-MIL approach addresses the challenges (discussed in section \ref{challenges}) in  CWS vs. CWNS affective states difference classification from physiological sensing signals.
\par
\subsubsection{Weakly Labeled Data and Multiple Instance Learning (MIL):} We employ a multiple instance learning (MIL) paradigm \cite{xu2012context,xu2012multiple,salekin2018weakly} to address the absence of fine-grain annotations in our data. In MIL, each input of a classifier is considered as a bag of instances $B = \{ x_1,x_2,...x_K\}$. Each bag $B$ has an associated single binary label $Y \in \{ 0,1 \}$ known during training. However, the labels of instances within a bag, i.e., $y_1 … y_K$ and $y_k \in \{ 0,1 \}$ for $k = 1\dots K$ are unknown.
As per conventional instance-based MIL assumption \cite{foulds2010review}, a positive bag has a label: $Y=0$ and a negative bag has a label: $Y=1$. A negative bag has at least one negative instance, and may contain positive instances (i.e., $ \exists x_j \in B,\ y_j = 1$). However, a positive bag contains positive instances only (i.e., $ \forall x_j \in B,\ y_j = 0$). Thus, the relationship between bag label $Y$ and instance label $y_j$ is: $Y=max_{i=1,..k}(y_i)$. 
\par
In this paper, a negative bag is the extracted features set from a CWS’s 20s physiological data, whereas a positive bag is from a CWNS. Features extracted from each 2s segment discussed in section \ref{feature_extraction} constitute an instance, and the collection of all instances of a 20s physiological sensing data constitute a bag. According to the instance-based MIL assumption, if the MIL classifier identifies that at least one instance is negative (2s segment, conveying CWSs' distinctive physiological response pattern), the 20s data would be detected as a CWS's response. In contrast, if none of the instances is identified as negative, the 20s data would be detected as a CWNS's response.
% identify the CWSs' subtle but distinctive response-patterns present in the physiological features which are sparsely spread across the signal. 
\par
\subsubsection{Temporal Dependency and Attention-MIL:}
Previous studies \cite{walsh2019physiological,azarbarzin2014relationship} have shown that the physiological response in arousal comprises temporal patterns. A limitation of instance-based MIL \cite{foulds2010review} is that it considers that the distinctive patterns (indicative to CWS or negative class) are sparse and independent. Moreover, it classifies a bag (i.e., one input) based on the instance with the highest likelihood of being negative. Hence, a large portion of the data remains unutilized, negatively affecting the classification performance.
\par
To address this challenge, this paper adopts an attention-based MIL approach named attention-MIL \cite{ilse2018attention,wang2018revisiting}. Attention-MIL is capable of identifying sparse distinctive patterns from weakly labeled data, captures sequential traits, makes inferences from the aggregation of all instances in a bag, and is shown to achieve better classification performance \cite{wang2018revisiting}. In contrast to the instance-based MIL, it generates a score or attention weight for each instance ($x_i$) in a bag indicating the likelihood of the instance ($x_i$) conveying CWS indicative distinctive patterns. Weighted instances (preserving their temporal patterns) are aggregated through an attention-based pooling function (section \ref{attn-block}) to generate a bag representation, from which the classifier makes the inference. 
\par
\subsubsection{Modality-wise Distinctive Pattern Extraction:} This paper evaluates multiple modalities: HR, EDA, RSP-amp and RSP-rate (section \ref{feature_extraction}) to measure the participants’ physiological response. Section \ref{challenges}'s observation demonstrates that arousal indicative sparse patterns do not simultaneously emerge in each modality. 
Such observation motivates the need for CWSs’ distinctive sparse patterns extraction independently from each modality. Hence, this paper applies the attention-MIL approach to each modality separately and generates modality-specific bag representations. %encompassing the CWS’s distinctive responses.
\par 
\subsubsection{Capturing Cross-modality Dependency:} As discussed in section \ref{challenges}, cross-modality relationships can be effective attributes in differentiating CWS's situational physiological response from CWNS. Hence, the presented approach must capture the cross-modality relationships. 
\par 
As discussed above, the modality-wise attention-MIL mechanism generates an independent representation for each modality.
To capture the cross-modality relations, the presented approach uses a novel \textit{modality fusion mechanism} which learns the pair-wise and unary relationships between each modality-embedding.

\subsection{Modality Invariant-MIL (MI-MIL) Approach}\label{MI-MIL}
\begin{figure}[h]
  \centering
  \includegraphics[height=2.3in,width=\textwidth]{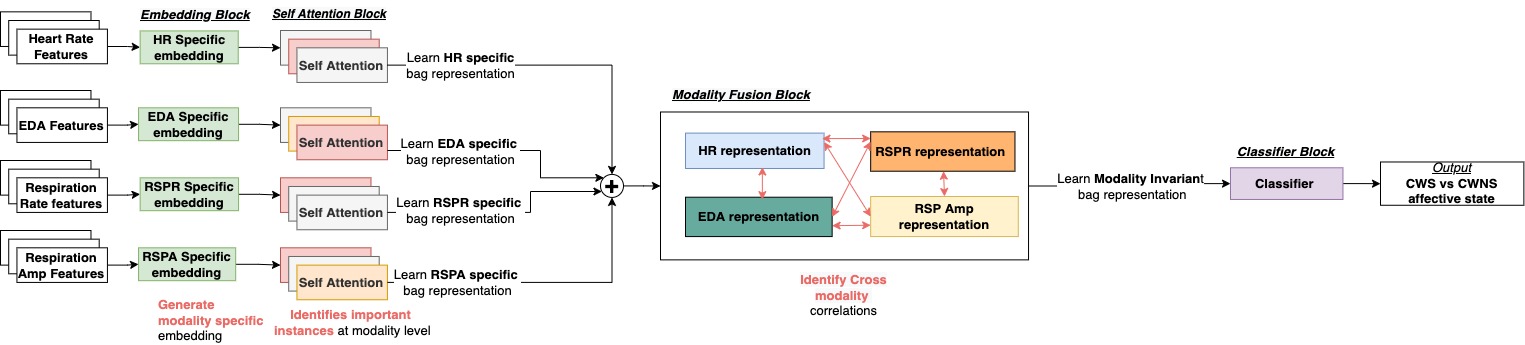}
  \caption{Modality-Invariant Multiple Instance Learning (MI-MIL). In this example figure, $k=3$, and four modalities $m=$EDA, HR, RSP-amp and RSP-rate. Each modality-specific input bag representation $B_m$ comprises three instances $x_{im}, i=1,2,3$. Modality specific embedding blocks generates three embeddings $e_{im}, i=1,2,3$. Each self-attention pooling block generates attention weights $a_{im}$ for the respective modality instance embeddings $e_{im}$. Weight values are shown with color, where darker color represents higher weight values. Each self-attention pooling block generates a modality-specific bag representation $t_m$ using a weighted average of the modality-specific embeddings. The modality fusion block takes the $<t_{EDA}, t_{HR}, t_{RSP-amp}, t_{RSP-rate}>$ vector as input and generates a modality invariant  representation $Z$, conveying the cross-modailty relations. A classifier takes the $Z$ as input and infers the class label (CWS vs. CWNS).}
    \label{MI-MIL archi}
    \vskip -3ex
\end{figure}

This section discusses the MI-MIL approach that takes the modality-specific bag representations ($B_m = \{ x_{1m},x_{2m}, \dots x_{km}\}$, $k=19, m=$ EDA, HR, RSP-amp, RSP-rate) of a 20s physiological sensing data as input. %Here, each instance $x_i$ is represented by the features extracted from each 2s contiguous segments (discussed in section \ref{feature_extraction}). 
As shown in figure \ref{MI-MIL archi}, MI-MIL has four components: (1) modality specific embedding block, (2) modality specific self-attention pooling block, (3) modality fusion Block, and (4) classifier Block. While the first two blocks are applied to each modality $m$ independently, the latter two combine the cross-modality information to generate inference. 
The components are discussed in detail below.

\subsubsection{Modality Specific Embedding Block:} For each modality $m$, the MI-MIL utilizes a modality-specific embedding block $f_m$ that takes each 2s segment (i.e., modality-specific instance $x_{im}$) of the respective modality as a separate input and transforms it into a lower $p$ dimensional embedding vector ($e_{im}$) \cite{golinko2019generalized}. Each block $f_m$ comprises multiple linear layers with ReLU activation functions. Generated embeddings $e_{im}, i=1,2…,k$ convey the modality $m$ specific CWS vs. CWNS differentiating information from their respective instance $x_{im}$.
%For $k=19$ input instances from the 20s detection window, $f_m$ generates $e_{im}, i=1,2…,k$ embeddings conveying the modality $m$ specific CWS vs. CWNS differentiating information from the respective instance $x_{im}$.
%The MI-MIL model first learns modality-specific features using the embedding block which consist of 3 Linear layers [24,256,256] which learns features that are specific to HR, EDA and Respiration features. Each features has a seperate embedding block and The features learned by the modality specific embeddings include signature patterns native to a single feature and information about the sequential and temporal dependencies of the input features at a modality level. 

\subsubsection{Modality Specific Self-Attention Pooling Block:} \label{attn-block} 
MI-MIL utilizes a modality specific self-attention pooling block for each modality $m$. It takes the modality specific embeddings $e_{im}, i=1,2,…,k$ preserving their temporal order as input, generates attention-weights $a_{im}$ for each of them, and generates a modality specific bag embedding $t_m$ following the equation \ref{attn_mech}.
\begin{equation}
\label{attn_mech}
    t_m=\Sigma^k_{i=1}a_{im}e_{im}\
    \ where:
    a_{im}=\frac{exp\{w_m^T tanh(V_m e_{im}^T)\}}{\Sigma^k_{j=1}exp\{w_m^T tanh(V_m e_{jm}^T)\}}
\end{equation}
Here, $w_m \in R^{L \times 1}$ and $V_m \in R^{L \times M}$ are the network parameters of the $m$ modality-specific self-attention pooling block. The hyperbolic tangent $tanh(.)$, element-wise non-linearity is utilized to ensure proper gradient flow \cite{ilse2018attention}. Generated weights $a_{im}$ represent the likelihood of the embedding $e_{im}$’s conveying CWS vs. CWNS differentiating pattern information, and the weights $a_{im}, i=1,2,...,k$ must sum to 1 to be invariant to the number of instances of a bag. Hence, this block ensures the temporal patterns present in each modality are captured, aggregate CWS vs. CWNS differentiating modality-specific information from each 2s segments (i.e., instances), and trainable through backpropagation.
\par 
As discussed in section \ref{Method-challenges}, CWS vs. CWNS differentiating patterns can be present in different asynchronous portions (i.e., different timestamps) of different modality-signals. Hence, different modality-specific pooling blocks may learn different attention weights for each 2s segment (i.e., instance), enabling CWS indicative pattern extraction independently from each modality. Equation \ref{attn_mech} is similar to the attention pooling mechanism presented in attention-MIL paper \cite{ilse2018attention}.
\par
\subsubsection{Modality Fusion Block:}
\label{modality-fusion}
Modality fusion block captures the cross-modality relationships.
It receives four independently generated modality-specific bag embeddings from four modality-specific self-attention pooling blocks. Each of the embeddings is an N-dimensional vector. 
It concatenates the four embeddings to a [1,4N] dimensional vector, $X=<x_1, x_2, … x_{4N}>$, and generates a [1,4N] dimensional vector, $Z=<z_1, z_2, … z_{4N}>$ that encodes the pairwise relations among all possible vector dimensions $(x_i, x_j)$ of $X$, as well as the unary relations, meaning how one dimension $x_i$ of $X$ may have its independent impact. Hence, $Z$ essentially encodes pair-wise and unary relationships between each of the dimensions of each of the modalities. Each $z_i$ encodes the relations corresponding to $x_i$ and computed using equation \ref{non-local-eqn}.

\begin{equation}
\label{non-local-eqn}
    z_i = \frac{1}{C(x)}\sum_{\forall{j}} f(x_i,x_j)g(x_j)
\end{equation}

Here, $z_i$ computation enumerates all possible dimensions $j$. $f(x_{i},x_{j})$ represent the \textit{pairwise} relation between dimension $i$ and $j$ of $X$. Here we use Embedded Gaussian function as function f:
$ f(x_{i},x_{j})=e^{\theta (x_{i})^{T} \phi (x_{j})}$.
\par
Here, $\theta(x_{i}) =W_{\theta}x_{i}$ and $\phi (x_{j})=W_{\phi}x_{j}$ are embeddings of $x_i$ \& $x_j$, and $W_{\theta}$ \& $W_{\phi}$ are learnable network parameters. In our implementation, $W_{\theta}$ \& $W_{\phi}$ are single-convolutional-layers with kernel size of $1$. $\theta (x_{i})^{T} \phi (x_{j})$ is the dot-product similarity. The normalization factor is set as $C(x)=\sum_{\forall j}f(x_{i},x_{j})$. With the equation above, $\frac{1}{C(x)}f(x_{i},x_{j})$ become a softmax operation along the dimension j. 
\par
The function $g(x_j)$ generates an unary embedding of $x_j$. It is a simple linear embedding:
$g(x_{j})=W_{g}x_{j}$, where $W_{g}$ is a learnable network parameter. In our implementation, $W_{g}$ is a single-convolutional-layer with a kernel size of $1$. Hence, according to equation \ref{non-local-eqn}, the modality fusion block generates a modality invariant  representation $Z$, encoding pair-wise relations between each of the modalities while preserving each modality's unary information.

\subsubsection{Classification Block:}
The classification block predicts the bag label (CWS vs. CWNS from 20s data), taking the modality invariant representation $Z$ as input. Our implementation uses two fully connected linear layers followed by a Sigmoid activation as the classification block.

\section{Experiments and Evaluation of MI-MIL}\label{EVAL}
This section evaluates the performance of our MI-MIL approach and different approach components.
We compared MI-MIL's performance with state-of-the-art attention-based-MIL approach \cite{ilse2018attention}, DNN CNN, LSTM, and LSTM with attention approaches. The architectural information for MI-MIL, attention-based MIL and the mentioned baseline models are discussed in  Appendix \ref{MI-MIL-param}, \ref{Att-MIL-IMPL} \& \ref{baseline-IMPL} respectively. The presented network parameter configurations were optimized by performing a grid search of the possible network-parameters. 
\par 
Following we present the dataset splits and evaluation metrics used for our evaluations. Later, section \ref{Classifier-Development-EVAL} discusses the models’ performance on the research question-wise tasks discussed in section \ref{problem_statement} and the evaluation conclusions. Finally, section \ref{eval-approach-components} evaluates the raw features' importance in addressing the research questions and real-time executability of MI-MIL.
\par
\paragraph{Dataset Split and Evaluation Metrics: }\label{datasplit} 
For each of the evaluations, we followed \textit{the person-disjoint hold-out method} \cite{cawley2010over}. We divided each dataset into three person-disjoint train, validation, and test sets, randomly selecting an equal number of participants from the CWS and CWNS groups. The distribution was as follows: test set (all data from 3-CWS and 3-CWNS), validation set (all data from 3-CWS and 3-CWNS), and training set (rest of the data). The same training, validation, and test set distributions were used for all evaluations of a dataset. To reduce contingency and avoid overfitting, classifiers were trained (on training+validation set) and evaluated (on the test set) three times with different seed values, and the average results are reported in this paper.
Evaluation results are presented with the metrics: recall, precision, accuracy, F1-score, and specificity.%, and Area-Under-the-Receiver-Operating-Characteristic-Curve (ROC-AUC) \cite{ROC-AUC2}. The ROC-AUC value near 1 means that the model has a good measure of separability. Conversely, a poor model has an AUC near 0, which means it has the worst measure of separability.

\subsection{Investigating the Research Questions through Classification Evaluation}\label{Classifier-Development-EVAL}
We evaluated MI-MIL binary classifiers for each of the questions.
We trained different binary classification models using two sets of input features: raw physiological features (section \ref{Raw_physio_features}) and change-score features (section \ref{Change-score-features}). 

\subsubsection{Evaluation of Q1: Differentiating the CWS vs. CWNS from Scripted `window-1’ Signal:}\label{Res-Q1-Eval}
To get insights for our research question \ref{q1}, we evaluated the models to differentiate the CWS vs. CWNS from ‘window-1’ of the scripted dataset. \textit{ ‘Window-1’ particularly comprises the participants’ physiological responses upon exposure to external stressors (negatively-valenced picture).} 
Tables \ref{qnone_raw} \& \ref{qnone_cs} show our evaluation results.
%Our evaluation results using raw and change-score features are shown in tables \ref{qnone_raw} \& \ref{qnone_cs}.
%\par
%\textcolor{red}{
%All the reported results for the Attention-MIL architecture in both scripted and free-speech dataset were trained and evaluated using different randomly selected seed values and the average results have been reported in the following sections.
%For the evaluation of the research question 1 and 2 we trained a model using data from both the windows in training and validation dataset. The performance of the model was evaluated separately for question-1 and question-2 by using test datassets which contained only window-1 samples and window-2 samples respectively. The distribution of participants remained the same as dicussed in \ref{datasplit} at all times. The intuition behind not training separately for window-1 and window-2 was to allow the model to get sufficient data samples while training process so it can distinguish between the physiological patterns between CWS and CWNS. Using a validation set with both windows was to assure that it performs equally well irrespective of encountering window-1 or window-2. During the testing phase we only provided window-1 data and window-2 data at a time and recorded the performance using our metric.   
%}
\begin{table}[ ]
\caption{Evaluation of Q1: CWS vs. CWNS from scripted dataset: `window-1’ signal}
\label{qnone_eval}
\begin{subtable}[]{0.495\linewidth}
\centering
\caption{Evaluation using raw features}
\label{qnone_raw}
\resizebox{\linewidth}{!}{%
\begin{tabular}{|c|c|c|c|c|c|}
\hline
 Model & Accuracy            & F1  & precision  & Recall       & specificity  \\ 
\hline
DNN & 0.58                     & 0.60        & 0.56        & 0.66                   &  0.50 \\ 
\hline
CNN  & 0.56                      & 0.62         &  0.54       & 0.72                    & 0.40  \\ 
\hline
LSTM & 0.69                  & 0.65        &  0.74       &0.59                     & 0.80  \\  
\hline
LSTM (with Attention) & 0.69                  & 0.65        &  0.74       & 0.59                    & 0.80  \\  
\hline
Attention-MIL & 0.86            & 0.85        & 0.92        & 0.81                  & 0.91  \\
\hline
MI-MIL & 0.88            & 0.90        & 0.83        & 0.98                    & 0.79  \\
\hline
\end{tabular}}

\end{subtable}
\hfill
\begin{subtable}[]{0.495\linewidth}
\centering
\caption{Evaluation using change-score features}
\label{qnone_cs}
\resizebox{\linewidth}{!}{%
\begin{tabular}{|c|c|c|c|c|c|}
\hline
 Model& Accuracy    & F1  &  precision \% & Recall & specificity  \\
\hline
DNN & 0.53        & 0.53          & 0.52        & 0.55  &  0.50 \\  
\hline
CNN & 0.54        & 0.54           &  0.53       & 0.55   & 0.53  \\  
\hline
LSTM & 0.53   &  0.52 &  0.52  &52    & 0.53  \\  
\hline
LSTM (with Attention) & 0.54   & 0.53  &  0.54  &  0.52   &  0.57 \\  
\hline
Attention-MIL & 0.63         & 0.69           & 0.59        & 0.85   & 0.42 \\  
\hline
MI-MIL & 0.80         & 0.80           & 0.79        & 0.82  &0.77 \\  
\hline
\end{tabular}%
}
\end{subtable}
\vskip -1ex
\end{table}
\par 
The MI-MIL approach achieves F1-scores of 0.90 and 0.80 (for raw and change score features respectively), establishing that CWS exhibit easily identifiable unique physiological `attributes' (raw features) and `fluctuation-from-neural-condition' (change-score features) patterns than the CWNS while exposed to external stressors. 
\par
Notably, MI-MIL outperforms all the baseline models.
Specifically, in table \ref{qnone_cs}, the lower performance of the Attention-based-MIL indicates that the existing CWS distinctive sparse situational fluctuation-from-neural-condition patterns are subtle and more disjoint across different modalities compared to the physiological attribute patterns. However, MI-MIL's modality-specific embedding blocks identify the patterns effectively, resulting in higher performance.
Supervised learning baseline approaches perform relatively poorly since they fail to optimize with the absence of fine-grain data annotations, hence confirming the need for a weakly supervised learning MIL paradigm for effective classification.

\subsubsection{Evaluation of Q2: Differentiating the CWS vs. CWNS from Scripted `window-2’ Signal:}\label{Res-Q2-Eval}
To get insights for research question \ref{q2}, we evaluated the models to differentiate the CWS vs. CWNS from `window-2’ of the scripted dataset. In `window-2’ of the scripted dataset, participants are repeating the predetermined BBAP phrase. \textit{Hence, this window comprises the participants’ physiological responses while talking under stressful conditions.} Our evaluation results using raw and change-score features are shown in tables \ref{qntwo_raw} \& \ref{qntwo_cs}.

\begin{table}[ ]
\caption{Evaluation of Q2: CWS vs. CWNS from scripted dataset: `window-2’ signal}
\label{qntwo_eval}
\begin{subtable}[]{0.495\linewidth}
\centering
\caption{Evaluation using raw features}
\label{qntwo_raw} 
\resizebox{\linewidth}{!}{%
\begin{tabular}{|c|c|c|c|c|c|}
\hline
 Model & Accuracy   & F1  & precision  & Recall  & specificity  \\
 \hline
DNN  & 0.63       & 0.67          & 0.59        & 0.76  &  0.50 \\  
\hline
CNN  & 0.61        & 0.69           &  0.57       & 0.90   & 0.33  \\  
\hline
LSTM & 0.80           & 0.79           &  0.81       &0.76    & 0.83  \\  
\hline
LSTM (with Attention) & 0.78   &  0.76 & 0.81   & 0.72    &  0.83 \\  
\hline
Attention-MIL & 0.88         & 0.84           & 0.91        & 0.80   & 0.91 \\ \hline
MI-MIL & 0.88          & 0.89           & 0.82        & 0.98   & 0.78 \\ \hline
\end{tabular}%
}
\vskip -5ex
\end{subtable}
\hfill
\begin{subtable}[]{0.495\linewidth}
\centering
\caption{Evaluation using change-score features}
\label{qntwo_cs}
\resizebox{\linewidth}{!}{%
\begin{tabular}{|c|c|c|c|c|c|}
\hline
 Model & Accuracy & F1 & precision & Recall  & specificity \\
\hline
DNN    & 0.58     & 0.55          & 0.58        & 0.52  &  0.63 \\  
\hline
CNN    & 0.58      & 0.56           &  0.57       & 0.55   & 0.60  \\  
\hline
LSTM   & 0.58      & 0.55           &  0.58       &0.52    & 0.63  \\  
\hline
LSTM (with Attention)  & 0.64 & 0.60  &  0.67  &  0.55   & 0.73  \\  
\hline
Attention-MIL   & 0.72 & 0.76           & 0.66        & 0.90   & 0.56   \\
\hline
MI-MIL   & 0.83 & 0.83           & 0.82        & 0.85  & 0.80   \\
\hline
\end{tabular}%
}
\vskip -5ex
\end{subtable}
\vskip -1ex
\end{table}
\par 
The MI-MIL approach achieves F1-scores of 0.89 and 0.83 for raw and change score features respectively, establishing that CWS exhibits unique physiological attributes (i.e., amplitudes) and fluctuation-from-neutral-condition patterns than the CWNS while talking in stressful conditions. 
%According to the table \ref{qntwo_raw}, the MI-MIL approach achieves a 89\% F1-score and 0.88 ROC-AUC using raw features, establishing that CWS depict identifiable unique physiological responses than the CWNS while talking in stressful conditions.
%\par
%Table \ref{qntwo_cs} shows our evaluations using change-score features. MI-MIL achieves a 83\% F1-score and 0.83 ROC-AUC, indicating the existence of differentiating patterns on how CWNs’ physiological parameter fluctuates differently than the CWNSs’ from their baseline neutral affective states. 
Comparing Table \ref{qnone_cs} \& \ref{qntwo_cs}, demonstrates that, the CWS's unique physiological parameters fluctuation patterns are more explicit and identifiable while talking in stressful conditions. MI-MIL significantly outperforms supervised learning baselines (i.e., CNN, DNN, LSTM, LSTM with attention) and MIL baseline (attention-based-MIL), demonstrating that the presence of sparse and modality-specific disjoint physiological response patterns in CWS while they speak under stressful condition.
%\par 
%Notably, the baseline approaches achieve low accuracy (F1-scores in the range of 67-76 and 55-60 in tables \ref{qntwo_raw} \& \ref{qntwo_cs}), indicating the existing patterns are sparse and independent (discussed in section \ref{challenges}); hence difficult to differentiate via supervised learning classifiers. The Attention-MIL on the other hand provides a boost as compared to the baseline models but is outperformed by the MI-MIL model. To understand why the MI-MIL outperforms the Attention-MIL we evaluated different network structures by performing an ablation study on the attention-mil architecture (\textcolor{red}{discussed in Appendix \ref{Ablation-eval}}), We empirically found that using modality specific embedding with Attention-MIL structure (modality specific embeddings with common attention mechanisms for all modalities) provided a 5\% boost in accuracy but the structure did not account for cross modality relationships.

\subsubsection{Evaluation of Q3: Differentiating the CWS vs. CWNS from Scripted Dataset Baseline Signal:}\label{Q3:Evaluationanalysis}
This evaluation addresses the research question \ref{q3}. It is important to understand if the CWS and CWNS show differences in physiology during their neutral affective state or baseline condition. If yes, the use of raw physiological features for classification may provide erroneous insights. For example, if CWS have higher physiological parameter values than the CWNS in baseline/neutral condition, identifying that the CWS and CWNS are showing similar parameter values (i.e., low classification accuracy) in a challenging situation would not demonstrate that they have similar physiological responses. Instead, it may indicate the CWNSs’ parameters fluctuation is higher (i.e., stronger physiological response) than the CWSs’. 
\begin{table}[]
\caption{Evaluation of Q3: CWS vs. CWNS from Scripted baseline signal}
\label{qnthree_raw}
\centering
\resizebox{0.5\textwidth}{!}{%
\begin{tabular}{|c|c|c|c|c|c|}
\hline
 Model & Accuracy            & F1 & precision  & Recall & specificity \\
 \hline
DNN   &  0.65     & 0.64          & 0.68        & 0.61  &  0.70 \\  
\hline
CNN    &  0.63    & 0.69           &  0.68       & 0.70   & 0.48  \\ 
\hline
LSTM    &  0.63    & 0.67           &  0.61       & 0.75    & 0.50  \\  
\hline
Attention-MIL &   0.58       & 0.65           & 0.65        & 0.73  & 0.43 \\ \hline
MI-MIL &     0.51     &  0.56          &    0.52     & 0.62    & 0.40 \\ \hline
\end{tabular}%
}
\vskip -2ex
\end{table}
\par 
Hence, we developed and evaluated models to differentiate CWS vs. CWNS from the 20s baseline signals of the scripted dataset. The evaluation results are shown in table \ref{qnthree_raw}. All models take raw physiological features as input and achieve low F1-scores (0.56-0.69). These results demonstrate that the CWS and CWNS exhibit similar physiological parameters during neutral/baseline conditions, though some subtle differentiating patterns exist. 
\par
Such findings justify our evaluation of the models using raw and change-score features separately. Models with raw features give us insights into the physiological parameters value differences. Models with change-score features give us insights into the fluctuations in physiology that represent stronger or weaker responses (i.e., higher or lower fluctuations).

\subsubsection{Evaluation of Q4: Differentiating the CWS vs. CWNS from Free-Speech Signal:}\label{Res-Q5-Eval}
Though no external stressors were imposed during the free-speech experiment, the narration task is linguistically and cognitively demanding. It may elicit different physiological responses in CWS vs. CWNS. To investigate such differences in responses, we evaluated the models to detect the differences in the CWS vs. CWNS from the free-speech dataset. Our evaluation results using raw and change-score features are shown in tables \ref{qnfive_raw} \& \ref{qnfive_cs}.
\begin{table}[]
\caption{ Evaluation of Q4: CWS vs. CWNS from Free-speech dataset}
\begin{subtable}[]{0.495\linewidth}
\caption{Evaluation using raw features}
\label{qnfive_raw}
\centering
\resizebox{\linewidth}{!}{%
\begin{tabular}{|c|c|c|c|c|c|}
\hline
 Model & Accuracy           & F1  & precision  & Recall & specificity  \\
 \hline
DNN &    0.43     & 0.60          & 0.44        & 0.93 &  0.01 \\  
\hline
CNN  &    0.55    & 0.66           &  0.50       & 0.94  & 0.23  \\  
\hline
LSTM  &    0.56    & 0.63           &  0.51       & 0.83    & 0.33  \\  
\hline
LSTM (with Attention) & 0.56  & 0.57  & 0.51    & 0.64  &  0.50 \\  
\hline
Attention-MIL &      0.70    & 0.69           & 0.66        & 0.74   & 0.67 \\ \hline
MI-MIL &      0.72    & 0.73           & 0.68        & 0.85   & 0.62 \\ \hline
\end{tabular}%
}

\end{subtable}
\hfill
\begin{subtable}[]{0.495\linewidth}
\caption{Evaluation using Change score features}
\label{qnfive_cs}
\centering
\resizebox{\linewidth}{!}{%
\begin{tabular}{|c|c|c|c|c|c|}
\hline
 Model  & Accuracy           & F1  & precision  & Recall& specificity  \\
\hline
DNN  &  0.55      & 0.42          & 0.50        & 0.36  &  0.71 \\  
\hline
CNN   &   0.50   & 0.43           &  0.44       & 0.41   & 0.57  \\  
\hline
LSTM   &  0.42      & 0.41           &  0.38       & 0.46  & 0.39  \\  
\hline
LSTM (with Attention) & 0.46   & 0.53  &  0.44  & 0.68   & 0.28 \\  
\hline
Attention-MIL &  0.66         & 0.64           & 0.61        & 0.71 & 0.62   \\
\hline
MI-MIL &  0.64         & 0.67           & 0.57        & 0.81   & 0.49   \\
\hline
\end{tabular}%
}

\end{subtable}

\vskip -1ex
\label{qnfive_eval}
\end{table}
\par
The MI-MIL approach achieves F1-scores of 0.73 and 0.67 for raw and change score features respectively, establishing that CWS exhibit identifiable unique physiological responses than the CWNS while performing linguistically and cognitively demanding tasks (i.e., narration). MI-MIL approach outperforms all the baseline models, demonstrating its higher efficacy in identifying subtle, sparse and modality-disjoint patterns.
\par 
Notably, no predetermined phrases or sequences of phonemes were uttered during the spontaneous narration task. Hence, differentiating CWS vs. CWNS was more challenging than the scripted dataset’s `BBAP' phrase repetition task. Therefore, this section’s lower evaluation accuracy (table \ref{qnfive_eval}) compared to the tables \ref{qnone_eval} \& \ref{qntwo_eval}, does not indicate that the CWS exhibit a more explicit or stronger physiological response difference than CWNS in the stressful talking task compared to the narration task.

\subsection{Investigating the Approach Components}\label{eval-approach-components}
This section evaluates the raw features' importance and real-time executability of MI-MIL.
\subsubsection{Feature Selection:}
\label{feature_selection}
This section evaluates the discriminative capability of the raw physiological features for the CWNS vs. CWS classification task. Notably, recent literature \cite{bach2011dynamic,nikolic2011measuring,tumanova2020emotional, tasko2007speech,schneider2003respiration} have established that the HR, EDA, RSP-rate, RSP-amp are effective features for affective states and stuttering individual’s physiological arousal detection. Hence, this section aims to identify the features' relative importance rather than feature selection.
Following prior studies \cite{wang2013feature,zhang2018supervised,deepa2021ai,guodong2020feature,zhang2018generalized}, we employ ridge-regression. Ridge regression adds `squared magnitude’ of coefficient as penalty term to the loss function, hence highly penalizes coefficient of less important features. Evaluations are done on the scripted dataset for a fair comparison, where all participants experienced a similar condition. Table \ref{feature_selection_ranks} presents the feature rankings. EDA features are highly important, following RSP-rate, HR, and RSP-amp features. These results are in line with our ML interpretability evaluations. 
\par
We also evaluate the performance of our baseline Attention MIL model utilizing the top-k features from the ranks, and the results are presented in Table \ref{top-k}. Notably, since not many modalities were selected during Top-5, -10 evaluations, MI-MIL was not utilized. According to our evaluation, adding more features result in higher performance. These evaluations indicate that our classifiers were not overfitting due to redundant features, which is obvious since we have only 24 features, and all of them are shown to be effective by the literature.
\begin{table}[h!]

\caption{\small Feature rankings based on coefficients of ridge classifier and classification performance utilizing top-$k$ features.}
\centering
\hfill
\begin{subtable}[]{0.49\linewidth}
\caption{Feature selection rankings}
\label{feature_selection_ranks}
\centering
\resizebox{0.99\textwidth}{!}{%
\begin{tabular}{|c|c|c|c|c|c|}
\hline
 Rank & Feature name      & Coefficient & Rank& Feature name & Coefficient  \\
 \hline
1 &   EDA Min       & 0.468    & 13 &    HR Median   &       -0.002\\ \hline
2 &   EDA Max      & 0.191     & 14 &    HR Min      &  -0.012      \\ \hline
3 &   EDA Max       & 0.159      & 15 &    HR Max      &  -0.017      \\ \hline
4 &   RSP rate Mean        & 0.151 & 16 &    HR Variance       &  -0.021      \\ \hline 
5 &   EDA Mean    & 0.057      &  17 &    RSP amp Median     &  -0.023     \\ \hline
6 &   EDA Variance        & 0.056   & 18 &    RSP amp STD      &   -0.045    \\ \hline
7 &   RSP rate Max       & 0.038     & 19 &    RSP rate STD     &   -0.048    \\ \hline
8 &   HR STD    &   0.031   & 20 &    RSP amp Mean      &    -0.065   \\ \hline
9 &   RSP amp Variance   & 0.065      & 21 &    RSP rate Min      &  -0.066    \\ \hline
10 & RSP amp Min      &   0.012    &  22 &    EDA STD      &   -0.079    \\ \hline
11 & HR Mean      &    0.008   & 23 &    RSP rate Median     &   -0.089   \\ \hline
12 &  RSP rate Variance      &     0.003 & 24 &    EDA Median     &  -0.202     \\ \hline

\end{tabular}%
}
\vskip -1ex
\end{subtable}
\hfill
\begin{subtable}[]{0.49\linewidth}
\caption{Performance of Attention MIL for Top-k features \centering (k=5,10,15,20)}
\label{top-k}
\resizebox{0.99\textwidth}{!}{%
\begin{tabular}{|c|c|c|c|c|c|}
\hline
 Model & Accuracy            & F1 & precision  & Recall  & specificity \\
 \hline
Attention-MIL (Top-5) &   0.77       & 0.79           & 0.72        & 0.89   & 0.66 \\ \hline
Attention-MIL (Top-10) &   0.83       & 0.84           & 0.77        & 0.91   & 0.75 \\ \hline
Attention-MIL (Top-15) &   0.84       & 0.82           & 0.95        & 0.72   & 0.96 \\ \hline
Attention-MIL (Top-20)       &   0.85       & 0.86           & 0.78        & 0.96  & 0.75 \\ \hline
\end{tabular}%
}
\end{subtable}
\vskip -2ex
\end{table}

\subsubsection{Execution Time and Resource Usage on Scalable Devices:}\label{real-time-eval}
We evaluated MI-MIL’s real-time executability and resource usage on scalable mobile devices: an Nvidia Jetson Nano and a smartphone. We run the MI-MIL models taking consecutive 20-sec windows for 10 minutes and record the running time and resource usage. According to the Tablse \ref{EDGE_EVAL}, MI-MIL takes $0.019$ sec and $0.005$ sec on Jetson Nano and smartphone to process each 20-sec physiological data window. The average CPU and GPU usages are also low. The results suggest that MI-MIL can perform real-time analysis on resource constraint devices.

\begin{table}[h!]
\caption{ Evaluation of MI-MIL on Jetson Nano and Google pixel 6. The Jetson nano is equipped with NVIDIA Maxwell GPU, Quad-core ARM processor, and 4GB memory. The smartphone is a google pixel 6, powered by Octa-core ( 2x2.80 GHz Cortex-X1 , 2x2.25 GHz Cortex-A76 and 4x1.80 GHz Cortex-A55) with 8GB memory.}
\label{EDGE_EVAL}
\centering
\begin{subtable}[]{0.495\linewidth}
\caption{ Evaluation of MI-MIL on Jetson Nano}
\label{jetson_nano}
\centering
\resizebox{0.995\textwidth}{!}{%
\begin{tabular}{|c|c|c|c|c|}
\hline
 Features   &    \begin{tabular}[c]{@{}c@{}}Scripted\\ raw\end{tabular}      & \begin{tabular}[c]{@{}c@{}}Scripted\\ change-score\end{tabular}  & \begin{tabular}[c]{@{}c@{}}Free speech\\ raw\end{tabular} & \begin{tabular}[c]{@{}c@{}}Free speech\\ change-score\end{tabular}  \\
 \hline
Average CPU usage  (\%)  & 10.90       & 11.13   &10.82  &10.3           \\  
\hline
Average GPU usage (\%)    & 0.07      & 0.108   & 0.124 & 0.09           \\  
\hline
\begin{tabular}[c]{@{}c@{}}  Average memory\\  usage (Mb)\end{tabular}     & 1703       & 1692   & 1688 & 1538       \\  
\hline
Run-time (s) & 0.017      & 0.018  & 0.018 & 0.021     \\ \hline
\end{tabular}%
}

\vskip -1ex
\end{subtable}
\hfill
\begin{subtable}[]{0.495\linewidth}
\caption{ Evaluation of MI-MIL on Google pixel 6}
\label{samsung}
\centering
\resizebox{0.995\textwidth}{!}{%
\begin{tabular}{|c|c|c|c|c|}
\hline
 Features   &    \begin{tabular}[c]{@{}c@{}}Scripted\\ raw\end{tabular}      & \begin{tabular}[c]{@{}c@{}}Scripted\\ change-score\end{tabular}  & \begin{tabular}[c]{@{}c@{}}Free speech\\ raw\end{tabular} & \begin{tabular}[c]{@{}c@{}}Free speech\\ change-score\end{tabular} \\
 \hline
Average CPU usage (\%)   &  2     & 4  & 6 &      6     \\  
\hline
\begin{tabular}[c]{@{}c@{}}Average memory\\  usage (Mb)\end{tabular}     &  166     & 169  &167 &    128    \\  
\hline
Run-time (s)  &  0.0061     & 0.0043  & 0.0058 &0.0055     \\ \hline
\end{tabular}%
}

\vskip -1ex
\end{subtable}

\vskip -1ex
\end{table}

\section{Interpretability Visualization and Discussion}
\label{interpretability}
MI-MIL’s inferences can be utilized to understand stuttering children’s psychological arousal during speaking from a \emph{group-wise} and \emph{personalized} perspective.
The developed binary ML classifiers in section \ref{Classifier-Development-EVAL} are BlackBox. Understanding which physiological features (raw or change-scores) are important for the respective model’s inference is critical since they contribute most to differentiating CWSs’ physiological response from others. We employed the KernelSHAP, a model-agnostic interpretation framework that determines each physiological feature’s (raw or change-score) impact in terms of its Shapley value \cite{shapley1953stochastic} on the respective model’s inferences. Notably, distinct Shapley feature importance values are generated for each input, indicating each feature's impact on generating a class inference for that particular input \emph{(i.e., personalized perspective)}.
%Shapley value, a measure from coalitional game theory, is the average marginal contribution of a feature value across all possible coalitions. KernelSHAP approximates the Shapley values for a model’s inference by defining a weighted least squares regression whose solution is the Shapley values for all the features.
%There is a simple principle of feature importance when evaluated using Shapley values. 
%The larger the absolute Shapley values, the more critical the feature is. In our model interpretations, positive Shapley values are responsible for moving prediction towards the CWNS class. In contrast, the negative Shapley values are responsible for moving the prediction towards the CWS class. For each 20s window that a model classifies, we get an array of Shapley values of the same dimension as the input; 19 $\times$ 24 or 19 $\times$ 8 for raw or change-score inputs. Each Shapley value of the 2D array represents the importance of the corresponding feature in the 2D-feature set (extracted from a 20s physiological sensing stream).
\par
However, Shapley values are additive \cite{lundberg2017unified}. We average the Shapley values per feature across the data to consider the \emph{global importance}. Global feature importance indicates how much a model relies on each feature at each timestamp, overall \emph{(i.e., group-wise perspective)}. We calculate the global importance for a model’s true predictions by computing the mean of the generated Shapley values corresponding to the test set 20s windows.
\par
\begin{figure*}
        \centering
        \begin{subfigure}[h!]{0.80\textwidth}
            \centering
            \includegraphics[width=\textwidth]{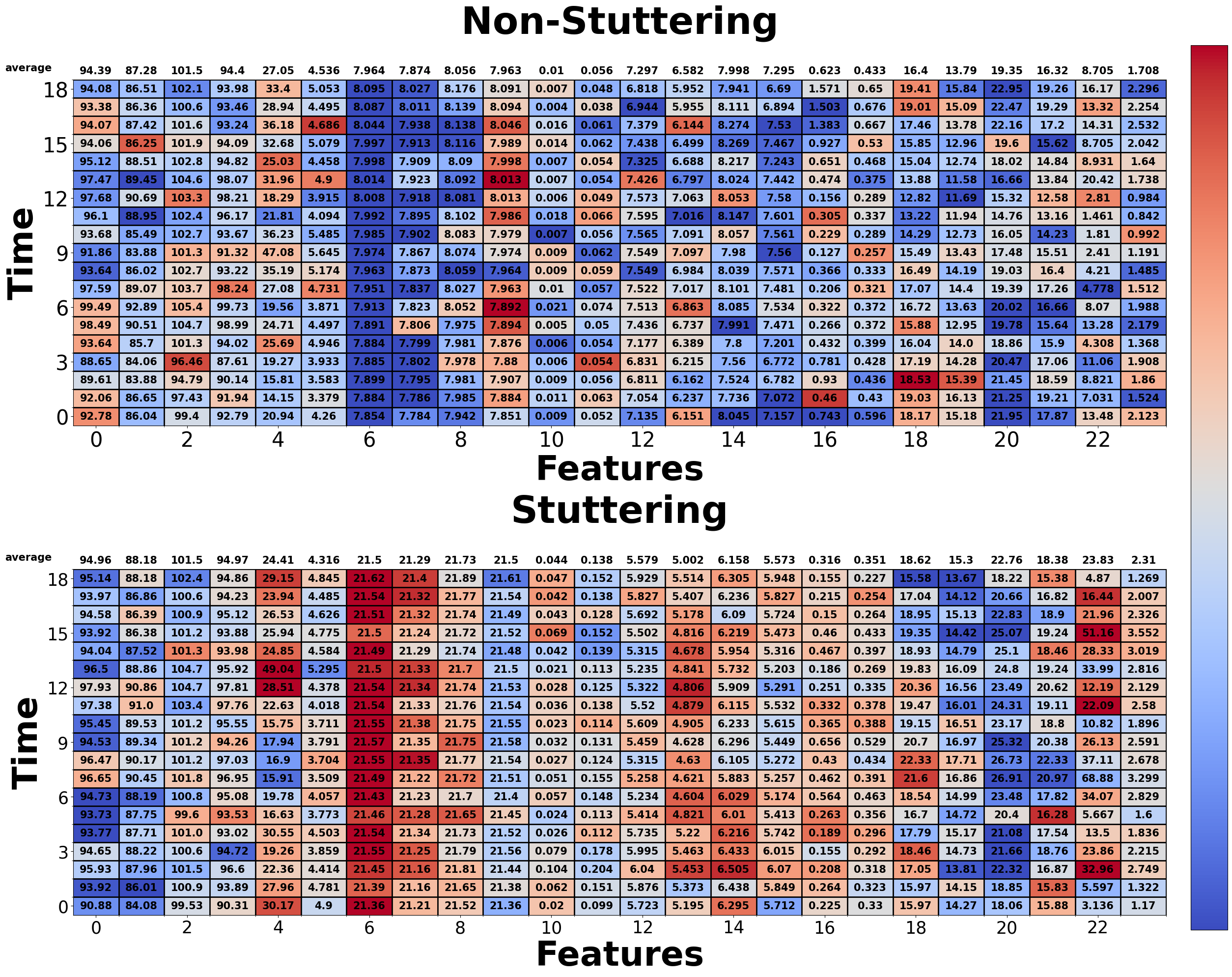}
            \vskip -1ex
            \caption[]%
            {{\small Window-2 Raw features}}    
            \label{fig:win-2_stu_raw}
        \end{subfigure}
        \hfill
        \vskip\baselineskip
        \vskip -2ex
        \begin{subfigure}[h!]{0.60\textwidth}   
            \centering 
            \includegraphics[width=\textwidth]{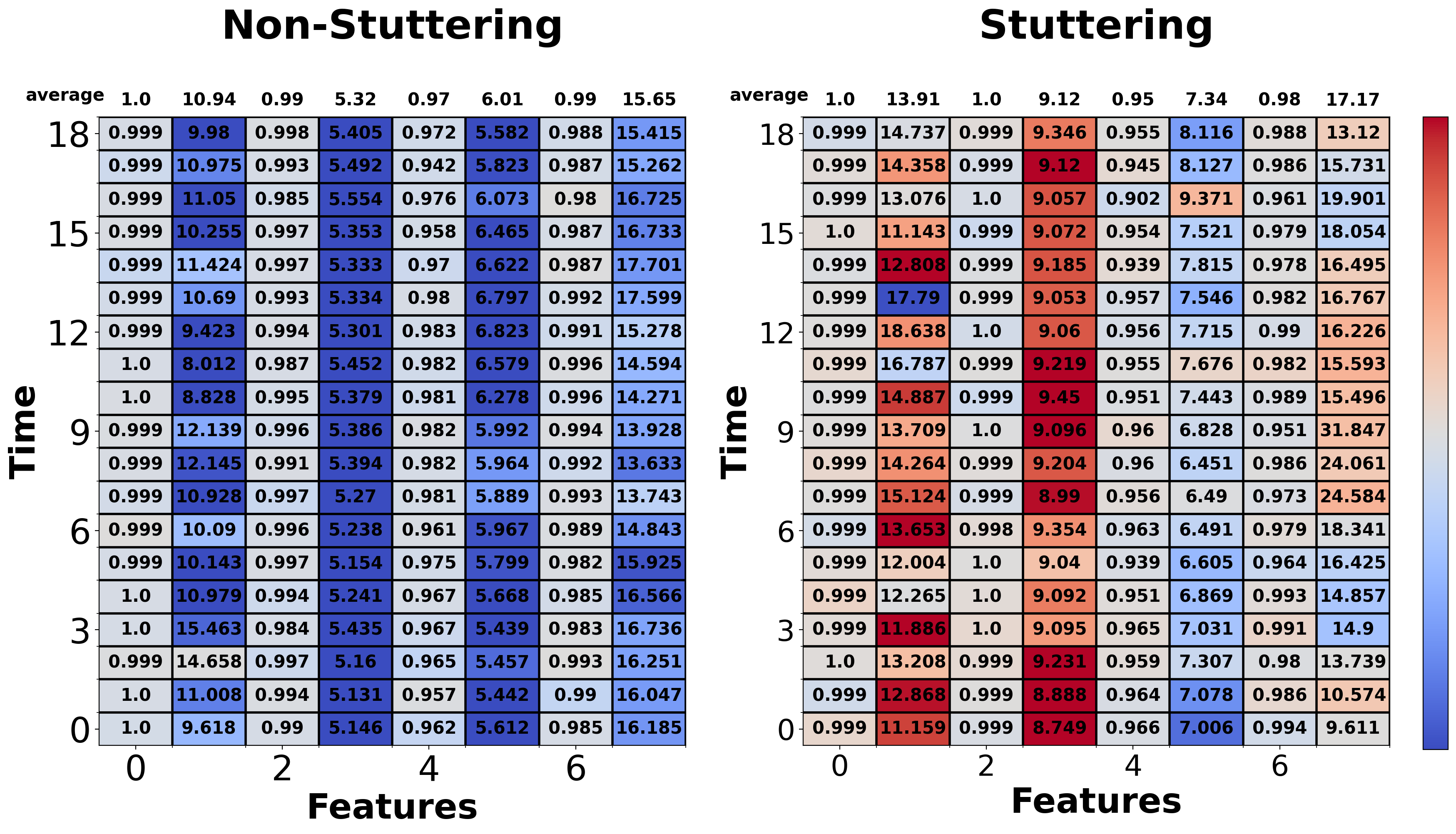}
            \vskip -1ex
            \caption[]%
            {{\small Window-2 change-score features}}    
            \label{win_2_stu_change}
        \end{subfigure}
        \hfill
        \vskip -1ex
        \caption[ ]%
        {\small Shapley global features-importance heatmaps for MI-MIL models evaluated on (Window-2) scripted dataset. In each plot, the X-axis represents the features (24 raw physiological and 8 change score features), and the Y-axis represents the ‘2s segments’ inside the 20s event detection window. Each cell in the heatmap contains the mean Shapley value of the respective feature in that 2s segment. The color-bars in the right of the figures show the colors scheme according to Shapley value’s magnitude. The raw and change-score features, with their corresponding indexes on the generated Shapley feature importance plots are shown in Appendix \ref{SHAP-feat-index} - table \ref{index_table}.} 
        \label{fig:win-2_graphs}
    \end{figure*}

We present the global feature importance for a research-question respective MI-MIL model’s inference through a grid-heatmap (with a cell for each features in each timestamp, e.g., figure \ref{fig:win-2_graphs}). In contrast, we present the personalized feature importance of each 20s physiological response data (from a CWS or CWNS) through time-series representation of the features and heatmap on each 2s segment of the time-series (section \ref{Q2-expln-personalize}).
\par
 The heatmaps use blue color to show positive Shapley values (responsible for pushing the model towards CWNS), and red color for negative Shapley values (responsible for pushing the model decision toward CWS), darker the intensity of the red/blue color higher the magnitude of the Shapley value and higher is the importance of the feature in pushing the model towards CWS/CWNS class. 
%The X-axis represents the features (24 raw physiological and 8 change score features). The Y-axis represents the ‘2s segments’ inside the 20s event detection window. Each cell in the heatmap contains the mean Shapley value of the respective feature in that 2s segment. 
%The heatmaps use blue color to show positive Shapley values (responsible for pushing the model towards CWNS), and red color for negative Shapley values (responsible for pushing the model decision toward CWS), darker the intensity of the red/blue color higher the magnitude of the Shapley value and higher is the importance of the feature in pushing the model towards CWS/CWNS class. The color-bars in the right of the figures show the colors scheme according to Shapley value’s magnitude.
\par
Speech science studies’ interest lies in understanding the second-to-second effect of children’s physiological arousal in their speech production; hence we are particularly interested in visualizing and understanding the importance of physiological features in differentiating CWS vs. CWNS while repeating the target phrase (BBAP) after viewing negatively-valenced (stress-provoking) pictures (discussed in section \ref{Q2-EXPL}), and while describing pictures spontaneously during the free-speech condition, discussed in Appendix \ref{Q4-EXPL}.

\begin{figure}[h!]
  \centering
  \includegraphics[width=0.7\textwidth]{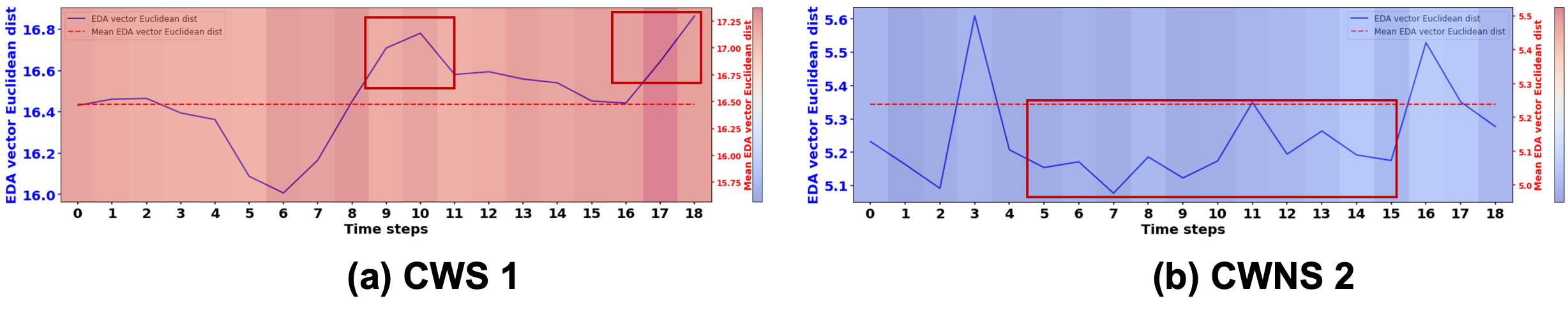}
    \vskip -3ex
 \caption{Participant wise shapley importance plots for EDA change score euclidean distance features (from one window-2)}
\label{personalized-EDA-SHAP}
\vskip -3ex
\end{figure}

\begin{figure}[h!]
  \centering
  \includegraphics[width=0.85\textwidth]{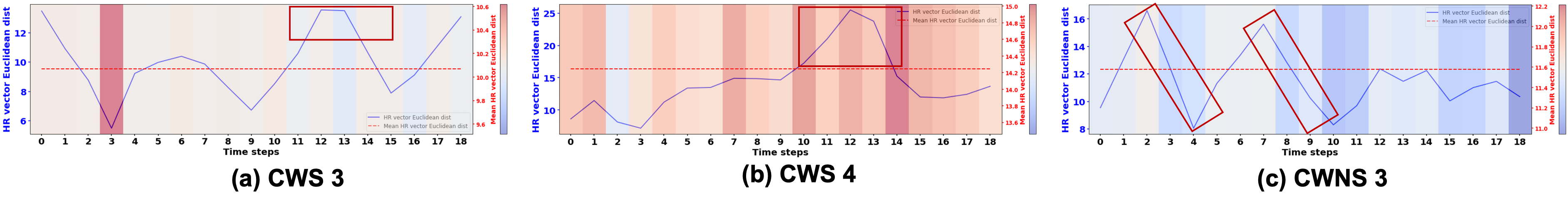}
  \vskip -3ex
 \caption{Participant wise shapley importance plots for HR change score euclidean distance features (from one window-2)}
\label{personalized-HR-SHAP}
\vskip -3ex
\end{figure}
\subsection{Q2 Interpretation: MI-MIL Model to Differentiate CWS vs. CWNS While Speaking in Stress Condition (Scripted Dataset)}
\label{Q2-EXPL}
This section discusses and demonstrates how MI-MIL’s inferences can be utilized and visualized to understand stuttering children’s distinctive psychological responses during speaking under stress condition from a group-wise (section \ref{Q2-expln-group}) and personalized (section \ref{Q2-expln-personalize}) level. The ML interpretations are discussed in following:

\subsubsection{Group-wise Global Feature Importance:}\label{Q2-expln-group}
The figures \ref{fig:win-2_stu_raw} and \ref{win_2_stu_change} show the global features importance through Shapley interpretation graphs of the MI-MIL models (section \ref{Res-Q2-Eval}) evaluated on window-2. 
Each of the figures comprises two SHAP plots: features’ importance plot for true negative (i.e., detecting CWS) and true positive classification (i.e., detecting CWNS). Window-2 comprises children’s physiology during the BBAP phrase repetition under stressful conditions task, and the evaluation in section \ref{Res-Q2-Eval} indicates that the CWS and CWNS show significant differences. This section visualizes and discusses the MI-MIL identified group-wise differences in CWS vs. CWNS through the respective model’s interpretation.
\par
\paragraph{EDA Features Importance:} As shown in figure \ref{fig:win-2_stu_raw}, CWS experienced higher raw EDA feature values than CWNS, indicating CWS experienced higher physiological arousal \cite{bach2011dynamic,nikolic2011measuring}. These raw EDA features are important in CWS vs. CWNS classification, indicating that they are distinctive patterns. Psychophysiological research examining speaking-related sympathetic nervous system activity in CWS is limited to just a few studies. However, speaking task- and age-related differences in skin conductance level have been reported \cite{choi2016emotional,tumanova2019autonomic,zengin2015sympathetic,zengin2018sympathetic,jones2014autonomic,jones2017executive,tumanova2020emotional}. \textit{This study is the first to evident that, the CWS experience distinctively higher arousal while talking in stressful condition than their non-stuttering peers.} When interpreting our raw EDA findings, the readers are reminded that in this evaluation, we used raw EDA data (phasic skin conductance responses were not removed from the signal; similarly, the baseline EDA was not considered in this model). 
\par 
\paragraph{HR Features Importance:}According to the figure \ref{fig:win-2_stu_raw}, for CWS participants, the HR variance during the later part of the 20s window (timesteps 12-18) shows a sudden increase and has high importance toward CWS classification (indicated by the dark red cells). Research indicates that social-emotional challenges or a feeling of anxiety elicit increased heart rate attributes in children and adults \cite{kudielka2004hpa}. Hence, \textit{we interpret the discussed HR-variance pattern as a sign that during talking under stressful conditions, CWS’s arousal increases with time, and it is a distinctive attribute between CWS and CWNS.}
Such interpretation is in line with the findings from the stuttering research literature. A recent statistical analysis-based study \cite{tumanova2020emotional} showed that CWS experienced on avg. higher HR attributes, hence, higher emotional arousal than CWNS. \emph{In contrast, our approach allows finding second by second temporal patterns in specific modalities like HR, which are distinctive in CWS.}
\par
\paragraph{RSP Features Importance:} In regard to the raw respiratory effort data (per the figure \ref{win_2_stu_change}), RSP-rate mean and variance contributed significantly to differentiating CWS vs. CWNS. Although the raw respiratory effort values were higher in CWS than in CWNS, the data for both groups are in line with speech breathing rates for preschool-age children \cite{boliek2009refinement}. 
Thus, \textit{our finding indicates the relatively faster speech breathing rates of CWS (compared to CWNS) as a distinctive respiratory effort pattern.} Interestingly, CWSs’ RSP feature values slightly decrease with time progression (in Y-axis), \textit{indicating that with the progression of time, CWS’s speech breathing rate decreases even under stressful conditions}.
Though, it has been proposed that stuttering is associated with various airflow irregularities (e.g., \cite{bloodstein2021handbook}), but to our knowledge, this is the first study to examine respiratory effort in preschool-age children who stutter. 
\par 
\paragraph{Change-score Features Importance:} We also evaluated the change-score of the four physiological features (HR, EDA, RSP-amp, and RSP-rate). According to the figures \ref{win_2_stu_change}, CWS showed a higher fluctuation from baseline in HR, EDA, RSP-amplitude, and RSP-rate features compared to CWNS. HR and EDA change score features have higher shapley importance. Hence, CWS showing higher fluctuation in EDA and HR during talking under stressful condition compared to the CWNS are distinctive patterns. It demonstrates that CWS experience higher arousal than their peers while speaking in stressful conditions.
\par
According to this section’s evaluation and interpretation, we can conclude that the CWS showed distinctive temporal and overall-window-wide physiological parameter differences from their non-stuttering peers. This paper's presented approach can identify these patterns in fine-grain, second by second level.

\subsubsection{Personalized Interpretation:}\label{Q2-expln-personalize}
On average, across CWS participants, the HR and EDA changes-scores from window-2 are distinctive. A major contribution of this paper is its ability to extract and visualize personalized fine-grain second by second temporal physiological response patterns. Figures \ref{personalized-EDA-SHAP} and \ref{personalized-HR-SHAP} demonstrate the visualization of personalized EDA and HR feature importance in 20s-window-2 from five different individuals. According to figure \ref{personalized-EDA-SHAP}, `CWS 1’ participant has higher EDA, meaning experiencing higher arousal \cite{bach2011dynamic,nikolic2011measuring} than the `CWNS 2’ participant. Moreover, the `CWS 1’ participant’s arousal increases with time, indicating talking in stressful situations in enhancing this participant’s arousal more and more. Notably, the MI-MIL Shapley feature importance values are higher (darker `red’) on those increasing  EDA picks, indicating the MI-MIL approach can effectively identify the personalized patterns. 
\par
According to figure \ref{personalized-HR-SHAP}, both CWS participants’ HR change-score is increasing with time, which is in line with section \ref{Q2-expln-group}. For `CWS 4’, the HR-change-score pick is a high value (timestamp 10-14) and has high Shapley feature importance values, similar to the CWS group. However, `CWS 3’ individual’s response is different. `CWS 3’ experienced a `freezing response’ at the beginning of talking (timestamp 3); hence the overall HR change-score values are not as high as the other CWS. Notably, the MI-MIL Shapley feature importance values are higher (darker `red’) on timestamp $3$ (during the freezing response), indicating the MI-MIL approach can effectively identify the distinctive personalized patterns.
\par
Additionally, in figure \ref{personalized-HR-SHAP}(c), the `CWNS 3’ individual’s HR change-score is relatively lower, has decreasing trends where the Shapley importance values are higher, which indicates the distinctive pattern of `CWNS 3’ individual that indicates the participant is not from the stuttering group.
\par
\emph{This section’s discussion demonstrates that our approach can effectively identify fine-grain, personalized, distinctive temporal patterns from CWS and CWNS individuals.} Identifying such patterns would enable personalized understanding of stuttering development and potential just-in-time personalized interventions to mitigate the physiological responses that may affect a children’s speech.

% \begin{figure}[h]
%   \centering
%   \includegraphics[width=0.7\textwidth]{PictureEDASHAP.png}
%     \vskip -2ex
%  \caption{Participant wise shapley importance plots for EDA change score euclidean distance features (from one window-2)}
% \label{personalized-EDA-SHAP}
% \vskip -2ex
% \end{figure}

% \begin{figure}[h]
%   \centering
%   \includegraphics[width=0.85\textwidth]{PictureHRSHAP.png}
%   \vskip -2ex
%  \caption{Participant wise shapley importance plots for HR change score euclidean distance features (from one window-2)}
% \label{personalized-HR-SHAP}
% \vskip -3ex
% \end{figure}

\section{Study Observations, Impact, and Limitations}\label{Summary-discussion}
We developed a novel MI-MIL approach (section \ref{MI-MIL}) that addresses the challenges present in differentiating CWS vs. CWNS’s situational physiological arousal (section \ref{challenges}) from `weakly labeled’ data. MI-MIL’s high efficacy in addressing all research questions indicates its effectiveness. 
\par 
%The presented study extracts raw and change-score features from the physiological parameters. Raw features convey information about the speaker’s physiology, where the change-score captures the fluctuation of physiological response from one’s baseline/neutral condition. Hence, their combination gives a comprehensive understanding of one’s situational physiological response. 
Our evaluation visualization in section \ref{Q2-EXPL} demonstrates presented papers approach can effectively identify fine-grain, personalized, distinctive temporal patterns from CWS and CWNS, group-wise and personalized for each 20s window. Notably, many of the patterns in CWS's physiological response patterns are investigated for the first time. Many of them conform to the existing speech science literature showing the reliability of our approach's visualization.
The following discusses the presented approach's impact, generalizability, and limitations.
\par 
\paragraph{Impact and Application of the Study:} Approximately 5 percent of all children go through a period of stuttering, and 1 percent suffer from long-term stuttering \cite{stutterpopulation}. Speech production is a complex process that requires fast and precise coordination of respiration, voice production, lip, tongue, and jaw movements (among other speech articulators) while simultaneously processing cognitive-linguistic information. Social engagement, including regulating own emotions and responding appropriately to one’s communicative partner, is also inherent to spoken communication. Naturally, speech production can be affected by the speaker’s physiological arousal. Young children who have speech disorders, such as stuttering, are especially vulnerable to these influences. 
\par 
The fast nature of speech production calls for fine-grain, second-by-second assessment of any physiological response parameter of interest that can influence speech characteristics. To attain this goal, the presented study offers a new way to examine CWS’s physiological arousal data and has both \emph{group-level} and \emph{personalized-level} impacts on stuttering and individuals with stuttering.
\par
\textit{Group-wise Impact:} This study’s presented ML-based group-wise examination (discussed in section \ref{Q2-expln-group}) of attributes and fluctuations in physiological arousal during speaking could inform our understanding of the role of physiological arousal in the development of stuttering and explain the origins of its situational variability, one of the key and unexplained features of this condition. 
\par
\textit{Personalized-level Impact:} 
Recent literature \cite{kraft2019role,alm2014stuttering,eggers2013inhibitory,kefalianos2012early} suggests that CWS (compared with fluent peers) have increased difficulty in the regulation and/or adaptation of their behavioral and emotional responses to everyday scenarios
which lead to increased emotional reactivity to stressful stimuli. This paper’s evaluation not only confirms that, moreover, the presented approach can identify the fine-grain, second by second, temporal and personalized distinctive physiological response patterns of each CWS. 
\par 
Important to note that the physiological sensing parameters evaluated in this study are present in recent wearables. For example, same Biopac sensors utilized in this study (section \ref{study_design}) are present in Biopac wireless wearable system \cite{biopacmobile}); hence the MI-MIL models can be implemented and evaluated in wearables. According to our evaluation in section \ref{real-time-eval}, developed MI-MIL models are real-time executable in smart devices (e.g., smartphones).
Hence, the presented approach has the potential to be leveraged for remote, continuous, automated, and real-time assessment of stuttering children’s physiological responses. Such assessments can be used clinically to facilitate providing just-in-time emotion regulation strategies (e.g., biofeedback, mindfulness interventions \cite{plexico2011mindful}) that may lead to improvement in speech fluency/disfluency.

\paragraph{Generalizability of MI-MIL:}It is important to note that the MI-MIL approach is not limited to CWS’s physiological response. To demonstrate the generalizability, we evaluated the MI-MIL model on a benchmark dataset named WESAD dataset \cite{schmidt2018introducing} which contains the physiological response parameters (RSP, ECG, EDA) from $15$ participants for baseline and stress conditions. Notably, WESAD dataset is not weakly labeled; has fine-grain annotations. Still, the MI-MIL approach outperformed all the baselines: DNN, CNN, LSTM, and Attention-based MIL models by achieving an F1 score of $0.92$ (Detail in Appendix \ref{WESAD}), which shows its generalizability, robustness, and applicability in different domains.
\par
\paragraph{Study Limitations:} The study's limitation is that we analyzed data from only two conditions. Future work would benefit from sampling data from a broader range of situations to determine the models’ predictive validity boundaries. 
Additionally, in the future, our approach can be implemented and evaluated on wearables for the longitudinal understanding of CWS’s situational physiological arousal. Notably, this study's scope does not include an evaluation of CWS's speech. Multi-modal analysis of speech acoustics and physiological parameters can be a future research direction.

\section{Conclusion}\label{conclusion}
The presented first-of-its-kind study effectively identifies and visualizes the second-by-second temporal pattern differences in the physiological arousal of preschool-age CWS and CWNS while speaking perceptually fluently in two challenging conditions: speaking in stressful situations and narration. We collected physiological parameters data from 70 children in the two target conditions. However, our dataset and differentiating CWS from CWNS leveraging their physiological response has several challenges (section \ref{challenges}), which we address by developing a novel MI-MIL. MI-MIL applies a multiple-instance-learning paradigm on each modality independently, while through a cross-modality-fusion network, it effectively combines each modality’s sparse, latent attributes. MI-MIL is real-time executable and generalizable to other domains. The evaluation of this classifier addresses four critical research questions that align with state-of-the-art speech science studies’ interests. Later, we leverage SHAP classifier interpretations to visualize the salient and fine-grain physiological parameters unique to CWS. Finally, comprehensive evaluations are done on multiple datasets, presented framework, and several baselines that identified notable insights on CWSs’ physiological arousal during speech production.
\section*{ACKNOWLEDGMENTS}\label{ACKNOWLEDGMENTS}
This work was supported in part by NSF IIS SCH $\#$ $2124285$, NIH NIDCD $\#$ $R21DC018103$, and NIH NIDCD $\#$ $R01DC017476-S2$. We extend our sincere gratitude to the children and their caregivers whose participation made this project possible.

\bibliographystyle{ACM-Reference-Format}
\bibliography{main}

\section{Appendix}

\subsection{Presented Change-score Feature's Efficacy}\label{change-score-eval-sec}
This section compares our vector distance (i.e., cosine similarity and euclidean distance) based change-score feature’s efficacy with conventional subtraction-based change-score features utilized in literature \cite{chiu2015complexity,clifton2019correlation,jones2017executive}. 
The conventional change-score for a variable ($\Delta_f$) at timestamp $t$ is calculated as the difference of the variable’s value on that timestamp ($f_t$) from its average at the baseline condition: $\Delta_{f}=f_t - f_{baseline Mean}$. The evaluation results utilizing conventional change-score features ($\Delta_f$) of HR, EDA, RSP-amp, and RSP-rate are shown in Table \ref{delta}, where results are relatively poor. Compared to the conventional subtraction based change-score, cosine similarity and the euclidean distance are effective measures for vector differences (i.e., our features from each timestamp is vectors) \cite{vijaymeena2016survey,kryszkiewicz2014cosine,gomez2019self,wang2005euclidean}, capable of subtle difference assessment from multi-dimensional feature vectors, resulting in higher performances (Tables \ref{qnone_eval}, \ref{qntwo_eval}, and \ref{qnfive_eval}).

\begin{table}[h!]
\caption{Evaluation of conventional $\Delta$-based change score features}
\label{delta}
\centering
\resizebox{0.5\textwidth}{!}{%
\begin{tabular}{|c|c|c|c|c|c|}
\hline
 Model & Accuracy            & F1 & precision & Recall & specificity  \\
 \hline
MI-MIL (Window-1 scripted) &   0.50       & 0.61           & 0.50        & 0.79   & 0.33 \\ \hline
MI-MIL (Window-2 scripted) &   0.57       & 0.65           & 0.54        & 0.82   & 0.33 \\ \hline
MI-MIL (Free speech)       &   0.49       & 0.47           & 0.40        & 0.41   & 0.23 \\ \hline

\end{tabular}%
}
\end{table}

\subsection{Feature Indexes Used in Shapley Interpretability Visualization}\label{SHAP-feat-index}
The table \ref{index_table} lists the features and their respective indices.
\begin{table}[h!]
\caption{Feature indexes}
\label{index_table}
\begin{subtable}[]{0.495\linewidth}
    \caption{Raw physiological features}
    \label{tab:raw-features}
   \centering
    \resizebox{0.8\linewidth}{!}{%
    \begin{tabular}{|c|c|c|c|}
    \hline
         \textbf{Index} & \textbf{Feature name} & \textbf{Index} & \textbf{Feature name}  \\
         \hline
                    0    &  Heart rate Mean & 12    &  RSP amplitude Mean\\
                    \hline
                    1    &  Heart rate Min & 13    &  RSP amplitude Min\\
                    \hline
                    2    &  Heart rate Max& 14    &  RSP amplitude Max\\
                    \hline
                    3    &  Heart rate Median& 15    &  RSP amplitude Median\\
                    \hline
                    4    &  Heart rate Var& 16    &  RSP amplitude Var\\
                    \hline
                    5    &  Heart rate Std&17    &  RSP amplitude Std\\
                    \hline
                    6    &  EDA Mean&18    &  RSP rate Mean\\
                    \hline
                    7    &  EDA Min&19    &  RSP rate Min\\
                    \hline
                    8    &  EDA Max&20    &  RSP rate Max\\
                    \hline
                    9    &  EDA Median&21    &  RSP rate Median\\
                    \hline
                    10    &  EDA Var& 22    &  RSP rate Var\\
                    \hline
                    11    &  EDA Std&23    &  RSP rate Std\\
                    \hline

    \end{tabular}%
    }
\end{subtable}
\hfill
\begin{subtable}[]{0.495\linewidth}
    \caption{Change score features}
    \label{tab:change-score-features}
    \centering
    \resizebox{0.8\linewidth}{!}{%
    \begin{tabular}{|c|c|}
    \hline
         \textbf{Index} & \textbf{Feature name}  \\
         \hline
                    0    &  Heart rate vector Cosine similarity\\
                    \hline
                    1   &   Heart rate vector Euclidean distance\\
                    \hline
                    2    &  EDA vector Cosine similarity\\
                    \hline
                    3   &   EDA vector Euclidean distance\\
                    \hline
                    4    &  RSP amplitude vector Cosine similarity\\
                    \hline
                    5   &   RSP amplitude vector Euclidean distance\\
                    \hline
                    6    &  RSP rate vector Cosine similarity\\
                    \hline
                    7   &   RSP rate vector Euclidean distance\\
                    \hline
    \end{tabular}%
    }

\end{subtable}

\end{table}

\subsection{Extension of MI-MIL to Other Physiological Datasets }\label{WESAD}
WESAD dataset \cite{schmidt2018introducing} contains physiological parameter data (RSP, ECG,EDA) from 15 participants (mean age: 27.5 $\pm$ 2.4 years) for baseline and stress conditions collected using a chest-worn RespiBAN device\cite{schmidt2018introducing}. During the stress condition, the participants performed a Tier Social Stress Test (TSST) \cite{kirschbaum1993trier}. HR and Respiration features (RSP-rate and RSP-amp) were extracted from raw ECG and Respiration effort signals. The same set of 24 statistical quantifiers (similar to the ones discussed in section \ref{Raw_physio_features}) were extracted from 20s overlapping windows (15s overlap). The table \ref{WESAD} illustrates the performance of various models on the WESAD dataset. The MI-MIL approach outperformed the baseline DNN, CNN, LSTM, and Attention-based MIL models by achieving an F1 score of 0.92. Among the baseline models, the Attention-MIL achieves the second-highest F1 score of 0.91; other baseline models also achieve high accuracy (in the range of 0.85-0.90). These results also advocate the success of weakly supervised approaches as both Attention-MIL and MI-MIL perform the classification task with high accuracy.
\begin{table}[h!]
\caption{ Evaluation baseline models and MI-MIL on WESAD dataset}
\label{WESAD}
\centering
\resizebox{0.5\textwidth}{!}{%
\begin{tabular}{|c|c|c|c|c|c|c|}
\hline
 Model   & Accuracy \%         & F1 \% & precision \% & Recall\%& ROC-AUC & specificity \% \\
 \hline
DNN    & 85      & 83          &    74     & 94  &0.87    & 80   \\  
\hline
CNN     &   90 &           87 &  84     & 91  & 0.91  &  90  \\  
\hline
LSTM     & 88  & 84           &  82      &87  &0.88   &   89 \\  
\hline
LSTM (with attention)     & 89  & 86           &  84      &88  &0.89   &   90 \\  
\hline
Attention-MIL  &   91      &    91        & 91        &91  &0.91    & 91 \\ 
\hline
MI-MIL  &        92 &     92       &    98     & 87 &   0.92 & 98 \\ \hline
\end{tabular}%
}

\end{table}

\subsection{Q4 Interpretation: MI-MIL model to Differentiate CWS vs. CWNS During Spontaneous Narration: Free-speech Dataset}
\label{Q4-EXPL}
This section discusses and demonstrates how MI-MIL’s inferences can be utilized and visualized to understand stuttering children’s distinctive psychological responses during spontaneous narration. The interpretations and corresponding observations at a group level are discussed in the following section.

\subsubsection{Group-wise Global Feature Importance}\label{Q4-expln-group}
The figures \ref{fig:win-2_stu_raw} and \ref{win_2_stu_change} show the global features importance through Shapley interpretation graphs of the MI-MIL models (section \ref{Res-Q5-Eval}) evaluated on the free speech dataset. 
Each of the figures comprises two SHAP plots: features’ importance plot for true negative (i.e., detecting CWS) and true positive classification (i.e., detecting CWNS). The free speech data comprises children’s physiology during a spontaneous narration, and the evaluation in section \ref{Res-Q5-Eval} indicates that the CWS and CWNS show significant differences. This section visualizes and discusses the MI-MIL identified group-wise differences in CWS vs. CWNS through the respective model’s interpretation.
\par
\begin{figure*}
        \centering
        \begin{subfigure}[ ]{0.80\textwidth}
            \centering
            \includegraphics[width=\textwidth]{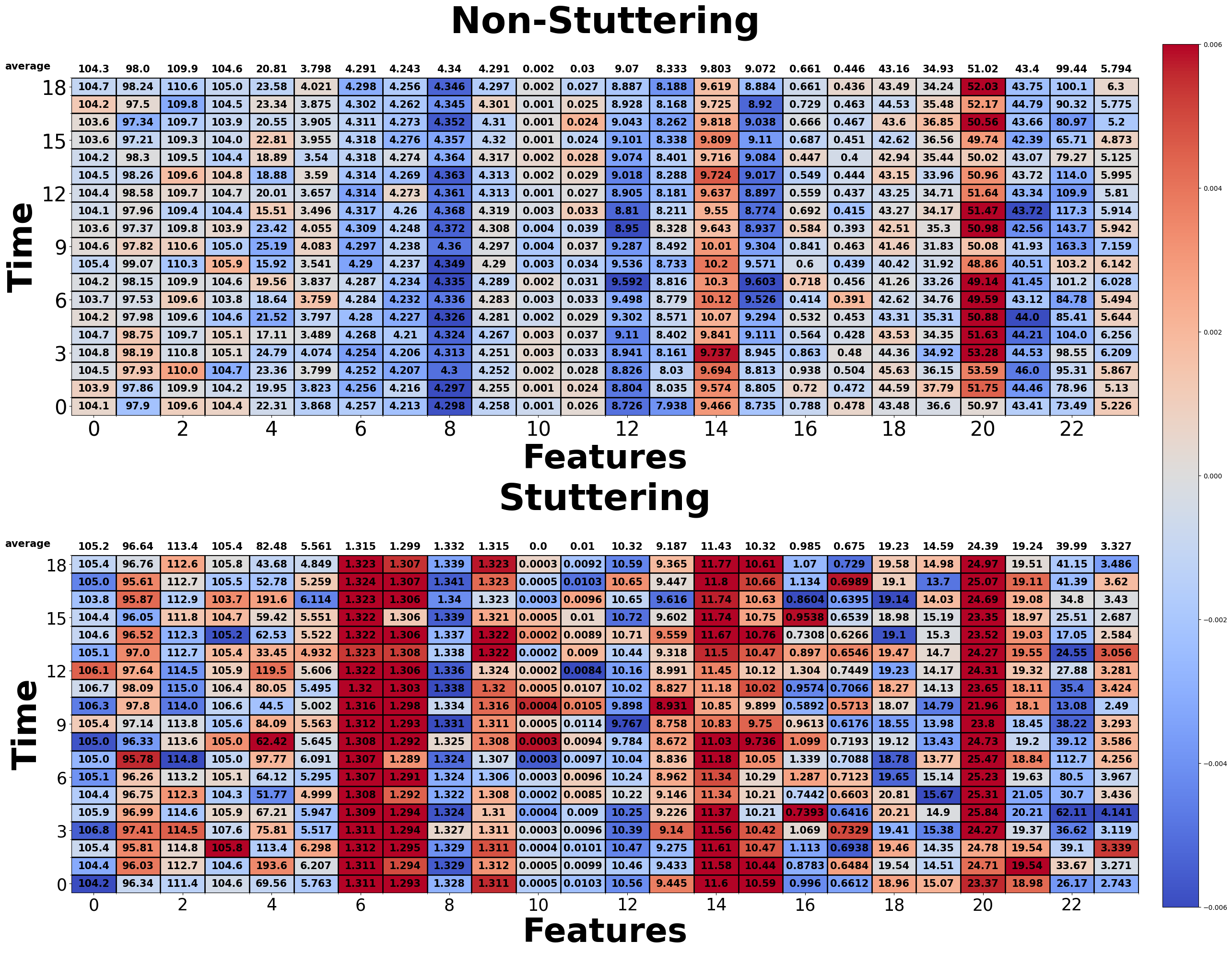}
            \caption[]%
            {{\small Free speech Raw features: shows the raw features’ importance plot for true positive classification (i.e., detecting CWS) and for true negative classification (i.e detecting CWNS).}}    
            \label{fig:fs_stu_raw}
        \end{subfigure}
        \vskip\baselineskip
        \begin{subfigure}[ ]{0.60\textwidth}   
            \centering 
            \includegraphics[width=\textwidth]{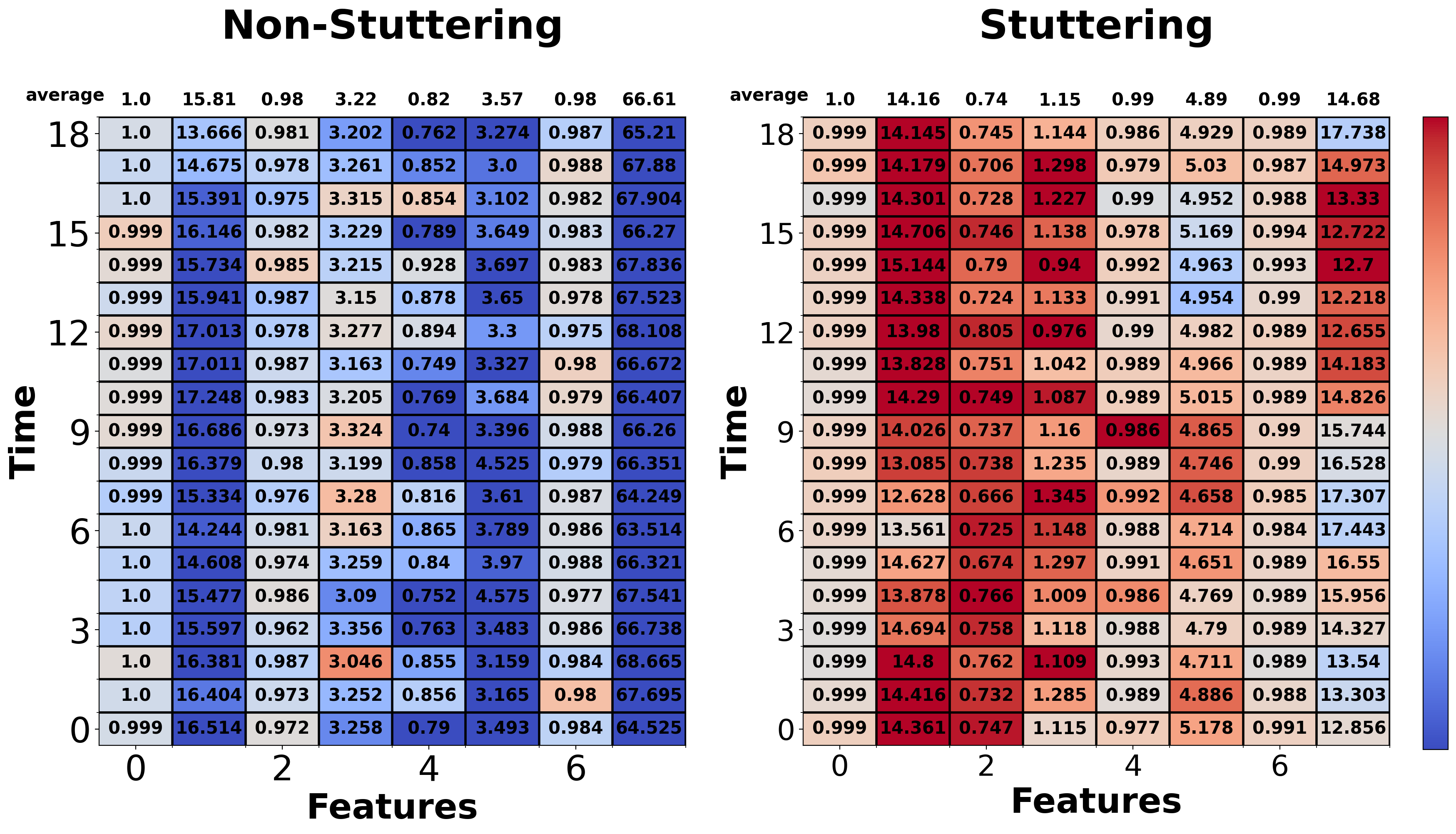}
            \caption[]%
            {{\small Free speech change score features}}    
            \label{fs_stu_change}
        \end{subfigure}
        \hfill
        \caption[ ]%
        {\small Shapley plots for models evaluated on free speech data} 
        \label{fig:fs_graphs}
    \end{figure*}

\paragraph{EDA Features Importance: }
As shown in figure \ref{fig:fs_stu_raw} both CWS and CWNS show low EDA feature values (indicating that they were not experiencing emotional stress or arousal. Specifically, the CWS show a lower EDA mean compared to CWNS thus EDA mean acts as an important differentiating feature for the classification task. Other important EDA features include EDA minimum and median which have high feature attribution (red blocks for stuttering class).
\par 
\paragraph{HR Features Importance: }
According to the figure \ref{fig:fs_stu_raw}, the only HR feature that differentiated CWS and CWNS during this task was HR minimum and variance. HR variance was significantly higher in CWS than in the CWNS. Both features were important feature in differentiating CWS (red grids in X-axis index 1 and 4) vs. CWNS (blue grids in X-axis index 1 and 4). However, other HR features were not different between the groups according to the ML model and pushed the classification towards the non-stuttering class. Thus, we interpret this data to suggest that there were certain time windows where the CWS showed a higher variation in the heart rate which the model recognised as important for the classification (dark red cells index 0-5).
\par
\paragraph{RSP Features Importance: }
CWNS showed slightly lower RSP-amplitude maximum and significantly higher RSP-rate maximum (feature index 20) compared to the CWS. The data for both groups are in line with speech breathing rates for preschool-age children \cite{boliek2009refinement}. It can be observed from the graph that the RSP-rate maximum has high contribution in classification of CWS class (red grids for feature 20). The difference in the magnitude of the of the RSP-rate and RSP-amp features helps that classifier in CWNS vs CWS classification (CWNS graphs have blue grids for feature 12,13,15 while CWS have red grids for feature 13,14,15,20). These results suggest that the RSP features have high contributions in the classification task for the spontaneous free speech task.  
\par 
\paragraph{Change-score Features Importance: }
The figure \ref{fs_stu_change} shows the change-score features’ importance plots. According to the figures, CWS showed lower increases from baseline in almost all physiologic features, namely HR, EDA, RSP-amplitude, and RSP-rate,   compared to those of CWNS. This suggests that CWNS experienced a higher arousal during the free speech task compared to CWS. The change score features were curated for the model to capture changes HR, EDA, RSP-amplitude and RSP-rate feature vectors from baseline in-terms of magnitude and direction. The euclidean distance and cosine similarity values of EDA, HR and RSP-rate feature vector were important for identifying CWS during the free speech task but these features contributed differently in different time steps as shown by the heatmap which are captured using the modality specific embedding and cross modality fusion architecture in the MI-MIL model.
\par
According to this section’s MI-MIL model interpretations for the free speech dataset, We observe that unlike the scripted dataset where CWS experienced a higher arousal during the scripted stress-provoking speech task compared to CWNS. For the spontaneous speaking task, CWNS showed higher physiologic values than CWS during this task. Lastly, both CWS and CWNS experienced higher arousal during the scripted stress-provoking speech task than during the free speech task.
.

\subsection{MIL}
\label{MIL_discussion}
This section discusses the 
 \textit{permutation-invariance property:} An implication that can be drawn from this relationship is that the order of the bag instances does not decide the corresponding probability of the bag. This assumption further motivates MIL use since the salient CWS affective state indicative signal patterns can be independent and sparse. 
\par 
\textit{MIL Decomposition:} The MIL model tries to predict the bag label using the equation \ref{mil_eqn}.
\begin{equation}
    \theta(X) =g(\sigma_{x\in X}{f(x)})
\label{mil_eqn}
\end{equation}
Here, $\theta(X) \in [0,1]$, $X$ is the set of instances in a bag, $f$ is the transformation function, $\sigma$ is the MIL pooling function, and $g$ is the classifier function. The choice of functions $f$, $g$, and $\sigma$ decide the specific approach to modeling the bag label probability. They are discussed below:

\begin{enumerate}
    \item \textit{Transformation function:} $f$ maps the instances ($x_j$,$j=1,\dots,k$) to low-dimensional instance-level representations ($e_j$) \cite{golinko2019generalized}. This function is applied on each individual instance belonging to the bag.
    
    \item \textit{MIL pooling function:} Symmetric function $\sigma$ takes the instance-level representations ($e_j$,$j=1,\dots,k$) as input to generate the bag-level representation. The $\sigma$ is hence also called the aggregation function. %Conventional MIL approaches consider the MIL pooling output as the predicted bag label.
    
    \item \textit{Classification function:} The function $g$ is only used in embedding-based MIL, discussed below. It takes the bag-level representation as input to generate a bag label probability score. It is used to pursue the final label of the bag. 
\end{enumerate} 
There are two variety of MIL approaches in literature \cite{carbonneau2018multiple,ilse2018attention}:
\par
\paragraph{Instance-Based MIL Approach:}\label{instance-based-approach} It is the common form of MIL evaluated in the literature. A \textit{transformation function $f$} takes each instance as input $x_j$ and returns the one-dimensional instance-level scores (i.e., instance labels $e_j=y_j$). Then individual instance-level scores are aggregated (through pooling) to obtain the bag label $Y$. Max or mean operators are generally used as the pooling functions. These two functions are symmetric and do not violate the permutation invariant assumption of the MIL approach. Studies have used other functions like convex maximum \cite{ramon2000multi,keeler1991integrated} as the pooling function $\sigma$. 
\par
\paragraph{Embedding-Based MIL Approach:} In this approach, a \textit{transformation function $f$} maps the instances ($x_j$,$j=1,\dots,k$) to a lower $m$-dimensional embeddings ($e_j$). $m$ is a hyper-parameter. A MIL pooling $\sigma$ takes the embeddings and generates a bag representation $z$ that is independent of the number of instances in the bag. A classifier function $g$ further process the bag representation to infer the bag label $Y$. This approach is arbitrarily flexible and can be trained by backpropagation. The only constraint is that the function $\sigma$ must be differentiable \cite{ilse2018attention,wang2018revisiting}. 
\par
Previous studies have shown that the embedding-based MIL approaches achieve better classification performance \cite{wang2018revisiting}. Since the individual instance labels are unknown, the transformation functions $f$ may be trained insufficiently and introduce error to the bag-level class prediction. The embedding-based approaches generate a joint representation of a bag from the instances; hence they do not introduce any additional bias to the bag-level classification \cite{ilse2018attention,wang2018revisiting}
\par 
\subsubsection{MI-MIL Implementation}
\label{MI-MIL-param}
\paragraph{As discussed in section \ref{Raw_physio_features}, $19\times24$ \textit{raw physiological features} are extracted from the 20s signal.} There are 19 instances, and each instance (i.e., 2s signal) is represented by a $1 \times 24$-dimensional feature set.
\par
\begin{figure}[h]
  \centering
  \includegraphics[width=\textwidth]{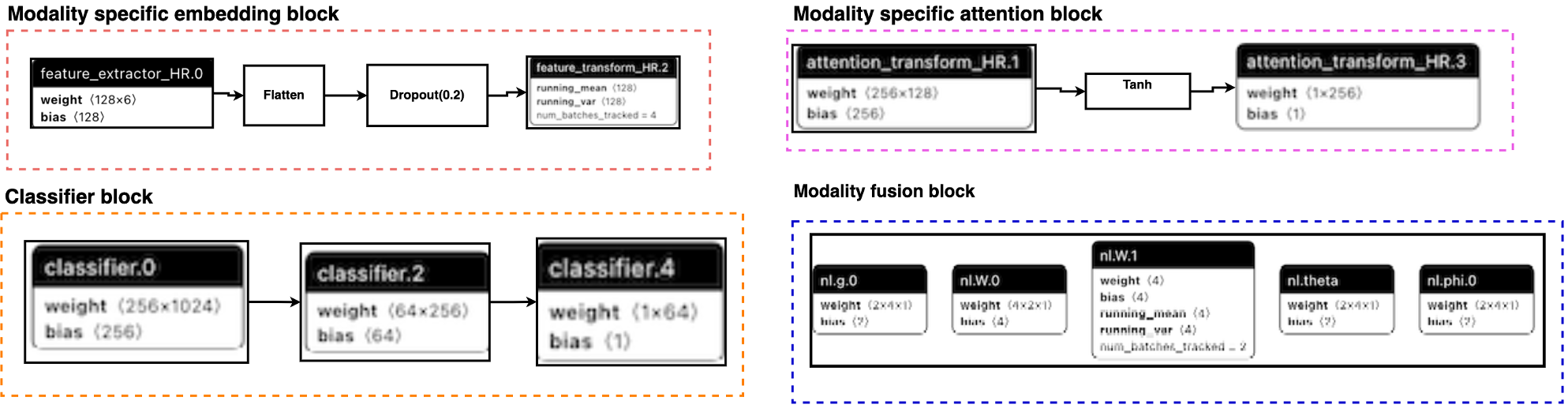}
  \caption{Layer-wise architectural information for MI-MIL}
    \label{MI-MIL_structure}
\end{figure}

\textit{The MI-MIL architecture for the  physiological features evaluated on the scripted dataset and free speech dataset is as follow:}
The MI-MIL implementation consisted of 4 separate modality specific embedding blocks. The embedding block for each modality consisted of a linear layer $[128]$ with a ReLU activation function. The output from the linear layer is flattened and passed through a  10 \% dropout layer which is followed by linear layer $[256]$. After generating a modality specific embedding we pass the embedding to the self attention block. The modality specific attention block consists of two linear layer$[256,1]$ which are separated by a $tanh$ function. The modality fusion block consists of three Conv1d layers with $4$ input channels and  $2$ output channels and a (kernel size=$1$,stride=$1$,padding=$0$). Since we concatenate embeddings from four modalities we get a combined bag representation of the dimensions <4,256> which is passed to the modality. The output from the modality fusion block goes through classification block which consists of three linear layers $[256,64,1]$ with ReLU activation between the first two layers and sigmoid activation function for the last layer of the classifier. The figure \ref{MI-MIL_structure} illustrates the layer wise structure for different blocks. 
\paragraph{We extract $19\times 8$ change-score physiological features from each 20s signal.} Thus, there are 19 instances, and each instance (i.e., 2s signal) is represented by a $1 \times 8$-dimensional change-score vector.
The Architecture of the change score MI-MIL model is same as the scripted dataset apart from the dropout layer (20\% dropout is used) in the HR and RSP rate modality embedding block.

\subsubsection{Attention-MIL Implementation}\label{Att-MIL-IMPL}

\textit{The attention-MIL architecture for the raw physiological features evaluated on the scripted dataset is as follow:}
The embedding block consists of a linear layer $[200]$, a 1-D convolution layer with $20$ kernel and $2X1$ kernel size, and another linear layer $[64]$ with ReLU activation function in each layer. A flatten layer is used to make the output two-dimensional, followed by a batch normalization layer and a linear layer $[128]$ with ReLU activation. The output of the embedding block is fed into the attention block consisting of two linear layer$[64,1]$ which are separated by a $tanh$ function (figure \ref{simple_attn}). The classification block consists of two linear layers $[64,1]$ with ReLU and sigmoid activation functions.
\par 
\textit{The attention-MIL architecture for the raw physiological features evaluated on the free-speech dataset is as follow:}
Since detecting affective states difference in CWS vs. CWNS from spontaneous narration is a harder task, and the free-speech data sample size is larger, the transformation/embedding block is more complex in this network. The transformation function block consists of four 1D-convolution layers $[256,512,128,32]$ and a kernel size of $2X1$ with ReLU activation. After flattening the output from the convolution layers, 1D batch normalization was used, followed by two linear layers $[512,128]$ with ReLU activations. The attention and classifier blocks were the same as the model evaluated on the scripted dataset discussed above.
\par
\textit{The attention-MIL architecture for the change-score features evaluated on the scripted dataset is as follow:}
The transformation function/embedding block has two 1-D convolution layers $[128, 64]$ followed by two linear layers $[256,256]$ and an intermediate 10\% dropout layer between the linear layers. Each of these layers uses a ReLU activation. Akin to the raw features model, the change score model flattens the output and feeds it into a 1D batch normalization, followed by a 10\% dropout layer and two linear layers $[512,256]$ with ReLU activation present after each of the layers. The attention block consisted of two linear layers $[64,1]$ with an intermediate $tanh$ function. The structure of the classification block consisted of four linear layers $[200,150,64,1]$. All linear layers had ReLU activation apart from the final layer, which had a sigmoid activation function.
\par 
\textit{The attention-MIL architecture for the change-score features evaluated on the free-speech dataset is as follow:}
The transformation function or embedding block has two convolution layers $[64,32]$ with a kernel size of $2X1$ trailed by two linear layers $[128,16]$. Each of the convolution and linear layers have a ReLU activation function. The output is flattened, fed into a 1D batch normalization layer, and fed into two linear layers $[1024,512]$. All the linear and convolution layers have ReLU activation functions. The attention block contains two linear layers $[64,1]$ separated by a $tanh$. The structure of the classification block remained the same as the raw features model evaluated on the scripted dataset discussed above.

\subsubsection{Baseline Model Implementation}\label{baseline-IMPL}
In each of the evaluations, we considered four supervised learning classifiers: long short-term memory (LSTM), convolutional network (CNN), deep neural network (DNN), and LSTM with self-attention network architectures as baseline models. The baseline model implementations are discussed below: 
\par 
\textit{LSTM model} was evaluated to explore the sequential dependence among the physiological data. The implementation consists of a single-layered LSTM and a linear output layer with Sigmoid activation. %We also evaluated the LSTM model with an attention mechanism where the attention block consisted of 2 linear layers $[250,1]$ separated by a $tanh$ function. A linear layer $[128]$ with ReLU activation and an output layer followed the attention block with Sigmoid activation. 
\par
The \textit{CNN model} can extract the complex sequential and global information of the data. It comprises 1D-convolution layers with kernel size $2X1$ and padding size of 1. A 20\% dropout is used between the convolution layers, and ReLU is used as the activation function. The output is then flattened and passed through some linear layers followed by 1D batch normalization layers. Each intermediate layer has ReLU as the activation function with Sigmoid activation in the output layer.
\par
The \textit{DNN model} architecture consists of some Linear layers, which are flattened and passed through the 1D batch normalization layer. After batch normalization, the outputs are passed through a dense layer with a sigmoid activation function.
\par
In order to see the performance of supervised learning approaches using the attention framework. We evaluated the \textit{LSTM model (with self attention)}. The model architecture consists of a four-layer LSTM whose output is fed to a self-attention block (similar to the one discussed in \ref{attn-block}) which consists of two linear layers separated by a $tanh$ activation. The attention block generates the attention weight for each instance after receiving the embedding from the LSTM layer. Then, the attention-weights are multiplied with the embedding generated from the LSTM model (self-attention), which is passed through a linear layer $[128]$ and an output layer.

\subsection{Overview of the existing deep learning literature on stress detection}\label{related-work-appendix-stress}
The table \ref{prior_work} lists the existing work in deep learning domain which are focused on stress detection. 
\begin{table}[h!]
\caption{Overview of the existing literature on Stress detection (Abbrev.- BVP:Blood Volume Pulse, BR: Breathing rate, EDA: Electrodermal activity,FDA:Fisher Discriminant
Analysis, GSR: Galvanic skin response,HR: Heart rate,PD: Pupil Diameter,RESP: Respiration activity, ST: Skin Temperature, SVM: Support Vector Machine, KNN: k- nearest neighbours, CNN: Convolutional Neural Network) }
\centering
\resizebox{0.75\textwidth}{!}{%
  \begin{tabular}{|p{0.3\textwidth}|p{0.15\textwidth}|p{0.15\textwidth}|p{0.15\textwidth}|p{0.3\textwidth}|}
    \toprule
    
    Stress Type & Sensor data & Literature  & Age-group&Method\\
    \hline
   Paced Stroop Test & GSR,
   BVP,
   ST,
   PD &\cite{zhai2006stress}&  21-42 & SVM\\
    \midrule
    Spielberger stress measurement questionnaire
    Memory test & EDA,
    HR & \cite{goumopoulos2019stress}&  59.8+-5.8 & Pvalue and Correlation analysis\\
    \midrule
    Talk Preparation,
    Hyperventilation & 
    HR,
    EDA &
    \cite{de2011stress}
    &
    19-32
    &
    Fuzzy Logic\\
    \midrule
    Mental Arithmetic Task, ,
    logical puzzle task & ECG,
    EDA,RESP
    EMG &
    \cite{wijsman2011towards}& 
    19-53 &
    Bayes Normal ,
    Quadratic Bayes Normal,
    K-NN ,
    Fisher's Least Square Linear Classifier \\
    \midrule
    Controlled Trier social stress test (TSST) &
    EDA,
    PPG,
    HR &
    \cite{mozos2017stress} &
    18-39 &
    SVM,AdaBoost,
    KNN\\
    \midrule
   Paced Stroop Test & EDA,
   BVP,
   ST,
   PD &\cite{zhai2008stress}&  21-42 & SVM
   Naives Bayes,
   Decision Tree\\
   \midrule
   Stroop Test &
   PPG,
   ECG,
   PD &\cite{6168563}& 22-28 & SVM
   Genetic Algorithm,
   Fuzzy SVM\\
   \midrule
   Deception  &
   GSR,
   BVP,
   ST,
   BR,
   &\cite{abouelenien2016human}&  20-35 & Decision Tree Classifier\\
   \midrule
    Talk Preparation,
    Hyperventilation & 
    HR,
    GSR &
    \cite{de2010two}
    &
    19-32
    &
    FDA, 
    KNN Classifier\\
    \midrule
    Applicants played custom version of game: tetris  & BVP,
    HR,
    EDA &
    \cite{maier2019deepflow}
    &
    18-32
    &
    CNN\\
    \midrule
    TSST  & RESP,
    ECG,
    EDA,
    EMG,
    ST &
    \cite{hssayeni2021multi}
    &
    25-30
    &
    CNN\\
    \midrule
    Scary Video, 
    Puzzle Task, 
    Pain, 
    Physical Activity  & EDA,
    BVP &
    \cite{albertetti2021stress}
    &
    25-30
    &
    Decision Tree,
    RNN,
    RCNN\\
    \midrule
  \bottomrule
\end{tabular}
}

\label{prior_work}
\vskip -6ex
\end{table}

\iffalse

\fi

\end{document}